\documentclass[superscriptaddress,twocolumn,pre]{revtex4-2}
\usepackage{amsmath}
\usepackage{amssymb}
\usepackage{graphicx}
\usepackage{hyperref}
\usepackage{amsfonts}
\usepackage{bbm}
\usepackage{tikz}

\usepackage{array}
\usepackage{cellspace}

\begin{document}
\title{The Role of Elastic Anisotropy in Active Nematics}
\author{Alexander J. H. Houston}
\email{Alexander.Houston@glasgow.ac.uk}
\affiliation{School of Mathematics and Statistics, University Place, Glasgow, G12 8QQ, United Kingdom.}

\begin{abstract}

We analyse the effect of anisotropy in elastic constants on the hydrodynamics of active nematics. Building on the multipole framework for a single elastic constant, we determine the leading effect of elastic anisotropy on the active response of generic distortions. The key findings are a new active torque, proportional to the anisotropy, in response to monopole distortions, and modifications to the propulsion of dipoles in both the direction of motion and changes in speed of up to 50\%. For point defects in two dimensions we find that, despite the large morphological changes in the director field, elastic anisotropy has only a minor impact on their hydrodynamics, with the self-propulsion speed of $+1/2$ defects lowered by less than 5\%. Finally, we determine the elastic torques exerted on defect pairs due to elastic anisotropy.
\end{abstract}

\maketitle

\section{Introduction} 

The dynamics of many living systems is determined not only by the motion of the individual agents, but also by the mechanical stresses that exist between them. This has been shown to be true for a variety of cells \cite{vining2017mechanical,ladoux2017mechanobiology,hayward2021tissue} and bacteria \cite{persat2015mechanical}, with impact on the phenomena of wound healing \cite{brugues2014forces}, extrusion \cite{eisenhoffer2012crowding}, jamming \cite{boocock2023interplay}, division \cite{gudipaty2017mechanical} and myoblast fusion \cite{le2024mechanical}. Understanding the mechanisms behind such processes, and how to use them to engineer desired behaviour in vitro \cite{zhang2021autonomous}, are central endeavours in the field of active matter.

Under the umbrella of active matter \cite{ramaswamy2010mechanics,marchetti2013hydrodynamics}, the hydrodynamics of active nematics \cite{doostmohammadi2018active} has been successfully applied to model bacteria \cite{wensink2012meso,zhou2014living}, cells \cite{duclos2017topological}, tissues \cite{saw2017topological,saw2018biological} and suspensions of synthetic microtubules \cite{sanchez2012spontaneous}. Here, the competition between mechanics and motility referred to above appears as a tension between elasticity and activity, the former seeking to maintain an ordered, quiescent state, and the later promoting a distorted, flowing one through fundamental hydrodynamic instabilities \cite{aditi2002hydrodynamic,ramaswamy2010mechanics}. This tension is manifest across the features of these systems, such as in the
spontaneous flow transitions exhibited under confinement \cite{voituriez2005spontaneous,marenduzzo2007steady,edwards2009spontaneous,duclos2018spontaneous}, their transition to active turbulence \cite{shendruk2017dancing,doostmohammadi2017onset}, and resultant statistical properties \cite{wensink2012meso,giomi2015geometry,thampi2016active,alert2020universal,alert2022active}.
Much effort has gone into controlling the behaviour of active nematics by influencing one of these factors, for example by varying forms of confinement \cite{norton2018insensitivity,opathalage2019self,duclos2017topological,chandrakar2020confinement}, patterning of activity \cite{zhang2021spatiotemporal,shankar2024design,partovifard2024controlling,houston2024spontaneous,schimming2025turbulence} or tuning of the elastic constants \cite{zhang2018interplay,kumar2018tunable,joshi2019interplay,kumar2022catapulting}.

It is common to model active nematics within the one-elastic-constant approximation, in which the elastic cost of splay, bend and twist deformations are all taken to be equal. While this affords simplification to both numerics and analytic calculations, it suppresses the geometric distinctiveness of the separate contributions to nematic elasticity. Even in passive nematics, elastic anisotropy can be the source of qualitative changes in behaviour and key to many phenomena~\cite{lavrentovich2024splay}. For active nematics, where the active force couples directly to splay and bend, the geometry and distinct elasticity are relevant {\sl a priori}. 
Extensile systems are hydrodynamically unstable to bend deformations \cite{ramaswamy2010mechanics,simha2002hydrodynamic}, softening the bend elastic constant and resulting in a large disparity in effective elastic constants \cite{kumar2018tunable}. This is reflected in the density of defects, which is determined by the bend elastic constant \cite{kumar2018tunable}, and in the defect morphology, which resembles the minimizer of a highly anisotropic elastic energy \cite{zhang2018interplay,kumar2018tunable,joshi2019interplay,kumar2022catapulting}. The competition between elastic and active stresses is especially pronounced for topological defects, since the large distortions they cause makes them epicentres of both. In particular, +1/2 defects behave as actively-motile quasiparticles \cite{giomi2013defect,giomi2014defect}, drive the transition to active turbulence \cite{shendruk2017dancing} and play an important role in cell extrusion \cite{saw2017topological}, morphogenesis \cite{maroudas2021topological} and bacterial layer formation \cite{copenhagen2021topological}.

In this paper we provide analytic calculations that shine light upon the connection between mechanics and dynamics in active nematics with unequal elastic constants. First, we determine the flows induced by generic distortions in an aligned far field, by extension of the active multipole framework~\cite{houston2023active} to include elastic anisotropy. The fundamental monopole acquires a modified angular dependence of quadrupolar structure, which produces rotational flows and novel active torques. As an illustration of the general consequences, we calculate the modification of the active flows for the UPenn dipole and find that the direction of self-propulsion can reverse when the bend constant is sufficiently larger than splay. For two-dimensional multipoles we show that the propulsion speed of dipoles can be altered by up to 50\% and the direction of propulsion by more than 10 degrees. 
We then consider the effect of elastic anisotropy on the active dynamics of topological defects. Strikingly, we find that the self-propulsion velocity is only affected at second order in the anisotropy with the result that the change in speed remains less than 5\%. Thus, one-elastic-constant calculations of defect speeds are reliable even when there is significant anisotropy. However, the anisotropy does give rise to an orientation-dependent term in the elastic interaction between defect pairs, which consequently experience a separation-independent torque due to elastic anisotropy. 

\section{Active Nematics}
We begin with a brief review of the description of active nematics \cite{marchetti2013hydrodynamics,doostmohammadi2018active} that we shall use in this paper, along with the active nematic multipole formulation \cite{houston2023active} that will aid us in characterising far-field distortions and establishing a correspondence between the geometry of the distortion and the active response. For our purposes, the hydrodynamics of active nematics is captured by their director field $\mathbf{n}$ and flow field $\mathbf{u}$. The former relaxes according to
\begin{equation}
    \partial_t\mathbf{n}+\mathbf{u}\cdot\nabla\mathbf{n}+\mathbf{\Omega}\cdot\mathbf{n}=\frac{1}{\gamma}\mathbf{h}-\nu\left[\mathbf{D}\cdot\mathbf{n}-\left(\mathbf{n}\cdot\mathbf{D}\mathbf{n}\right)\mathbf{n}\right],
\end{equation}
where $D_{ij} = \frac{1}{2} (\partial_i u_j + \partial_j u_i)$ and $\Omega_{ij} = \frac{1}{2} (\partial_i u_j - \partial_j u_i)$ are the symmetric and antisymmetric parts of the flow gradients, $\gamma$ is a rotational viscosity and $\nu$ is the flow alignment parameter. The molecular field $\mathbf{h}=-\delta F/\delta \mathbf{n}$ is the variational derivative of the Frank free energy, $F=\int f\text{d}V$, with the corresponding energy density given in three dimensions by
\begin{equation}
    f=\frac{1}{2} K_1\left(\nabla\cdot\mathbf{n}\right)^2+K_2\left(\mathbf{n}\cdot\nabla\times\mathbf{n}\right)^2+K_3\left(\mathbf{n}\cdot\nabla\mathbf{n}\right)^2,
\end{equation}
where $K_1$, $K_2$ and $K_3$ are the splay, twist and bend elastic constants. The same form holds in two dimensions, except that there is no twist ($K_2=0$). The flow is subject to mass continuity, $\nabla\cdot\mathbf{u}=0$, and the Stokes' equation, $\partial_i\sigma_{ij}=0$. The stress tensor is given by
\begin{equation}
    \begin{split}
        \sigma_{ij} & = - p \delta_{ij} + 2\mu D_{ij} + \frac{\nu}{2} \bigl( n_i h_j + h_i n_j \bigr) \\
        & \quad + \frac{1}{2} \bigl( n_i h_j - h_i n_j \bigr) - \frac{\partial f}{\partial (\partial_i n_k)} \partial_j n_k - \zeta n_i n_j ,
    \end{split}
\end{equation}
where $p$ is the pressure, $\mu$ an isotropic viscosity, $f$ the free energy density and $\zeta$ the activity, a phenomenological parameter which is positive in extensile active nematics and negative in contractile ones.

A simplification to the above description that we employ is to determine the active flow induced by an equilibrium director field. This permits analytical progress and has been fruitfully applied to the active flows around point defects \cite{giomi2014defect,angheluta2021role}, incorporating surface curvature \cite{khoromskaia2017vortex} or active turbulence \cite{alert2020universal}, three-dimensional defect loops \cite{binysh2019three,houston2022defect} and colloids \cite{houston2023active}, including intricately-shaped cogs \cite{houston2023colloids}. In this setting the director relaxation reduces to $\mathbf{h}=0$, with the Stokes' equation becoming
\begin{equation}
    -\nabla p+\mu\nabla^2\mathbf{u}=\zeta\nabla\cdot\left(\mathbf{n}\mathbf{n}\right).
    \label{eq:StokesActiveForce}
\end{equation}
Note that the Ericksen stress has been incorporated into a redefinition of the pressure, using the fact that $\mathbf{h}=0$ \cite{de1995physics}. We remark also that in this work we focus on the effects of elastic anisotropy, and so while we retain distinct elastic constants, we neglect the distinct viscosities that arise in nematics \cite{de1995physics} in favour of a single shear viscosity. In the discussion we suggest how the approach taken here could be extended to incorporate viscous anisotropy.

\subsection{Active nematic multipoles}
The response of active nematics to generic distortions has been characterised by the framework of active nematic multipoles \cite{houston2023active}. Provided the distortions have zero total topological charge, the far-field director consists of small harmonic deviations from uniform alignment. Explicitly, in three dimensions we write $\mathbf{n}=\mathbf{e}_z+\delta n_x\mathbf{e}_x+\delta n_y\mathbf{e}_y$, with both $\delta n_x$ and $\delta n_y$ harmonic. The same form holds in two dimensions, except that we find it convenient to take $\mathbf{e}_y$ as the far-field direction. The harmonic director components may be decomposed into harmonic multipoles, each expressible as a set of derivatives acting on a monopole, $1/r$ in three dimensions, or in two dimensions $\ln(r/R)$ for some large lengthscale $R$. At linear order the active response induced by a generic distortion may be found by applying the same set of derivatives to a fundamental flow or pressure response associated to the monopole, producing a hierarchy in which the monopole responses have an analogous status to the Stokeslet. A three-dimensional monopole in the $x$-direction may be written as $\mathbf{n}\approx\mathbf{e}_z+\alpha\frac{a}{r}\mathbf{e}_x$, with $a$ characterising the scale of the distortion and $\alpha$ a dimensionless coefficient. The associated fundamental flow response is \cite{houston2023active}
\begin{equation}
    \mathbf{u}=\frac{\alpha a\zeta}{4\mu}\left\lbrace\mathbf{e}_x\left[\frac{z}{r}+\frac{x^2z}{r^3}\right]+\mathbf{e}_y\frac{xyz}{r^3}+\mathbf{e}_z\left[\frac{x}{r}+\frac{xz^2}{r^3}\right]\right\rbrace.
    \label{eq:FundamentalActiveFlow3DIsotropic}
\end{equation}
The flow response for a $y$-monopole results from exchanging $x$ and $y$. In two dimensions we take the monopole director field to be $\mathbf{n}\approx\mathbf{e}_y+\alpha\frac{\log(r/R)}{\log(a/R)}\mathbf{e}_x$ and the fundamental flow response is given by
\begin{equation}
    \begin{split}
        \mathbf{u}&=\frac{\zeta\alpha}{8\mu\ln(\frac{a}{R})}\biggr[\frac{x^2-y^2}{r^2}\left(-y\mathbf{e}_x+x\mathbf{e}_y\right)\\
    &\qquad\qquad\qquad+2\ln\left(\frac{r}{R}\right)\left(y\mathbf{e}_x+x\mathbf{e}_y\right)\biggr].
    \end{split}
\end{equation}

The net active force, $\mathbf{f}$, and torque, $\boldsymbol{\tau}$, acting on a region $\Omega$ can be found via integrals of the active stress. In three dimensions this gives \cite{houston2023active}
\begin{widetext}
    \begin{align}
        \mathbf{f}&=\int_{\partial\Omega}\zeta\mathbf{n}\mathbf{n}\cdot\text{d}\mathbf{A}\approx\int_{S^2}\zeta\left\lbrace\mathbf{e}_x\frac{z\delta n_x}{r}+\mathbf{e}_y\frac{z\delta n_y}{r}+\mathbf{e}_z\frac{x\delta n_x+y\delta n_y}{r}\right\rbrace\text{d}A,\label{eq:ActiveForce}\\
        \boldsymbol{\tau}&=\int_{\partial\Omega}\mathbf{x}\times\zeta\mathbf{n}\mathbf{n}\cdot\text{d}\mathbf{A}\approx\int_{S^2}\zeta\left\lbrace \mathbf{e}_x\left[\frac{xy\delta n_x}{r}+\frac{(y^2-z^2)\delta n_y}{r}\right]+\mathbf{e}_y\left[\frac{(z^2-x^2)\delta n_x}{r}-\frac{xy\delta n_y}{r}\right]+\mathbf{e}_z\frac{z(-y\delta n_x+x\delta n_y)}{r}\right\rbrace\text{d}A,
        \label{eq:ActiveTorque}
    \end{align}
\end{widetext}
where the final expressions result from linearising the active stress and taking the region to be spherical. With these approximations, one can read off from \eqref{eq:ActiveForce} and \eqref{eq:ActiveTorque} which distortions produce a given active force or torque. Taken together, \eqref{eq:FundamentalActiveFlow3DIsotropic}, \eqref{eq:ActiveForce} and \eqref{eq:ActiveTorque} allow the determination of the active response and also the inverse problem: which distortions are needed to generate a desired propulsive or rotational behaviour in an active nematic. This last point is further elucidated by classifying the distortions according to their spin, or rotational symmetry, about the far-field direction, since only distortions whose spin is commensurate with that of a force or torque will be able to generate one \cite{houston2023active}.

\begin{figure*}
    \centering
    \begin{tikzpicture}[scale=1.6,>=stealth]
        \node[anchor=south west,inner sep=0] at (0,0)
{\includegraphics[width=1\linewidth, trim = 0 0 0 0, clip, angle = 0, origin = c]{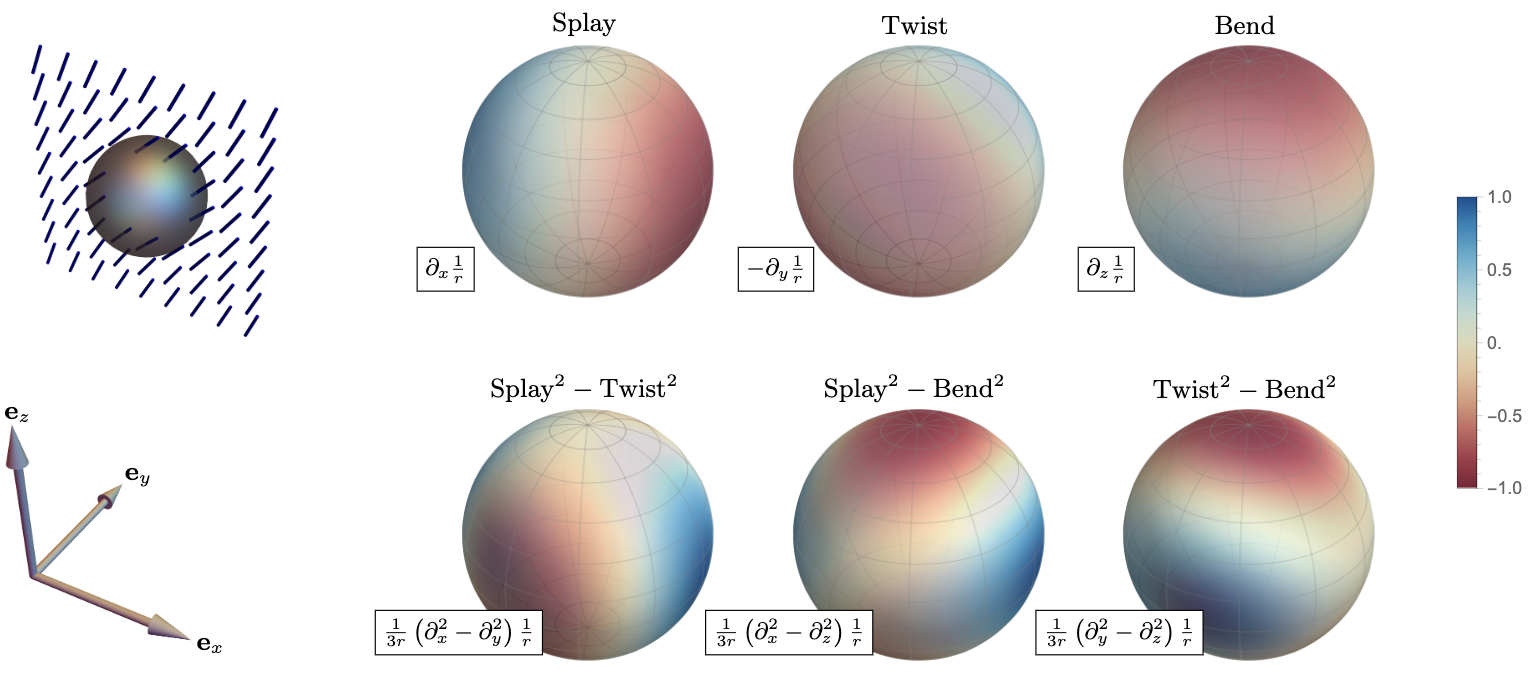}};
        
    \end{tikzpicture}
    \caption[Caption.]{The anisotropic distribution of elastic energy around a monopole. For a monopole distortion in the $x$-direction, shown top left, the distributions of splay, twist and bend is given by dipoles along $\mathbf{e}_x$, $-\mathbf{e}_y$ and $\mathbf{e}_z$ respectively. Unequal elastic constants therefore result in a quadrupole-like contribution to the elastic energy distribution, with the plane of the quadrupole depending on the type of anisotropy. This quadrupolar symmetry is inherited by the leading director perturbation, and hence is the root of many of the effects of elastic anisotropy in active nematics.
    }
    \label{fig:AnisotropicEnergy}
\end{figure*}

The spin describes the connection between local rotations of the director and global rotations of the distortion. For a spin-$s$ distortion, a uniform local rotation of the director about the far-field axis, $\mathbf{e}_z$ in three dimensions, corresponds to a global rotation scaled by a factor of $1/s$. Consequently, the magnitude of the spin corresponds to the rotational symmetry of the distortion about the far-field direction, with the sign indicating whether local rotations correspond to global rotations of the same or opposing sense. The exception is distortions with spin $0$, which have full rotational symmetry about the far-field direction, and for which local and global rotations are decoupled. The classification of distortions via spin is aided by an appropriate choice of derivatives: $\partial_z$, $\partial_{\bar{\omega}}=\frac{1}{2}(\partial_x+\text{i}\partial_y)$ and $\partial_{\omega}=\frac{1}{2}(\partial_x-\text{i}\partial_y)$, which maintain, lower and raise the spin respectively \cite{houston2023active}.

This framework has been used to determine the active dynamics of colloids, including the scalings with size of their linear and angular velocity \cite{houston2023active}, reveal spontaneous self-rotation due to buckling in active nematic defect loops \cite{houston2022defect} and, for cogs placed in active nematics, establish regimes of persistent spontaneous rotation and orientation-dependent chirality \cite{houston2023colloids}. In the following section we extend this formalism to include elastic anisotropy.

\section{Far-field distortions}
\label{sec:Far-field}
We begin our investigation of elastic anisotropy in active nematics by considering its impact on far-field distortions. We calculate the leading perturbation from spherical harmonics induced by elastic anisotropy by first calculating the molecular field. From this, and following the approach laid out in the previous section, we can calculate how elastic anisotropy modifies the active responses, forces and torques associated with these distortions.

Before embarking upon detailed calculations, we present a simple argument which underpins many of the results of this section. Considering a monopole distortion of the form $\mathbf{n}=\mathbf{e}_z+\alpha\frac{a}{r}\mathbf{e}_x$, we note that while, under a one-elastic-constant approximation, the elastic energy is isotropically distributed, the individual components of splay, twist and bend are not. Indeed, as illustrated in Figure \ref{fig:AnisotropicEnergy}, the leading contributions to splay, twist and bend correspond to orthogonal dipoles. Accordingly, anisotropy in the elastic constants leads to a component of the elastic energy distribution that has the symmetry, although not the radial decay, of a quadrupole. This quadrupole-like energy distribution, shown in the bottom row of Figure \ref{fig:AnisotropicEnergy}, is in the $x-y$-, $x-z$- or $y-z$-plane according to whether the anisotropy is of splay-twist, splay-bend or twist-bend type. As is typical in perturbation theory, and as we show in the following calculations, the leading correction to the director field inherits this quadrupolar symmetry from the energy distribution. From previous results on quadrupole-induced active torques \cite{houston2023active,houston2023colloids} we therefore expect monopole-like distortions in elastically anisotropic active nematics to result in active torques. Having inferred the effect of elastic anisotropy for the monopole, that for all distortions of higher order may be attained through differentiation. Although in what follows we shall find different combinations of the elastic constants more fruitful for exposing certain symmetries, the geometric argument given here is at the heart of the results we shall present.

\subsection{Three dimensions}
We start with three-dimensional distortions before presenting the two-dimensional case which, due to the lack of twist, is simpler. First, we establish the combinations of elastic constants that form a natural pair of dimensionless elastic anisotropy parameters, and then the leading order director perturbations to the monopole associated with each of these. One parameter captures the difference between the bend constant and the average of the splay and twist constants and induces a spin $1$ perturbation to the monopole. The other measures the difference between the splay and twist constants and results in a spin $-1$ perturbation. With these perturbations in hand we are able to determine the leading correction due to elastic anisotropy to the active flow response given in \eqref{eq:FundamentalActiveFlow3DIsotropic} and the net active force and torque, given in \eqref{eq:ActiveForce} and \eqref{eq:ActiveTorque} respectively. The most striking departure is, as alluded to above, that monopoles induce net active torques, whereas in the one-elastic-constant case they are free from both net active forces and torques.

\subsubsection{Morphology}
\begin{figure}
    \centering
    \begin{tikzpicture}[scale=1.6,>=stealth]
        \node[anchor=south west,inner sep=0] at (0,0)
{\includegraphics[width=1\linewidth, trim = 0 0 0 0, clip, angle = 0, origin = c]{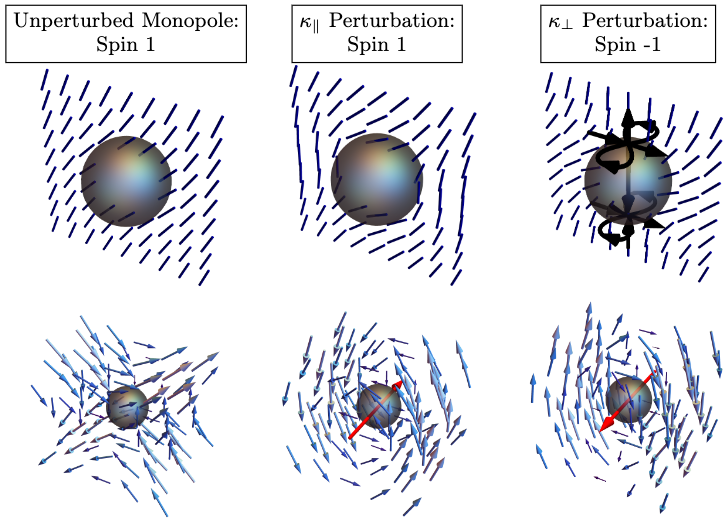}};
    \end{tikzpicture}
    \caption[Caption.]{The fundamental distortions and their associated active responses. The director field (above) and corresponding active flow (below) for, from left to right, the harmonic monopole, the perturbation due to parallel, or bend, anisotropy and the perturbation due to perpendicular, or splay-twist, anisotropy. The director field is shown as blue rods in a planar cross-section, with black arrows indicating its behaviour out of this plane where appropriate. The red arrows indicate the net active torque.
    }
    \label{fig:3DFundamentalDistortionsFlows}
\end{figure}

Working to linear order around a state that is, at least locally, uniformly aligned we may write the director field as $\textbf{n}=\textbf{e}_z+\delta n_x\textbf{e}_x+\delta n_y\textbf{e}_y$. 
Following the approach of \cite{houston2023active}, we complexify the plane transverse to the far-field direction, $\mathbf{e}_z$, introducing the complex director deformation $\delta n=\delta n_x+\text{i}\delta n_y$ and the complex derivatives $\partial_{\omega}=\frac{1}{2}(\partial_x-\text{i}\partial_y)$ and $\partial_{\bar{\omega}}=\frac{1}{2}(\partial_x+\text{i}\partial_y)$. 
Noting that $\partial_{\omega}\delta n=\frac{1}{2}[\nabla\cdot\textbf{n}+\text{i}\textbf{n}\cdot\nabla\times\textbf{n}]$ we write the Frank free energy as
\begin{widetext}
\begin{equation}
\begin{split}
    F &= \frac{1}{2}\int 4\left[K_1 \mathcal{R}\left\lbrace\partial_{\omega}\delta n\right\rbrace^2+K_2 \mathcal{I}\left\lbrace\partial_{\omega}\delta n\right\rbrace^2\right]+K_3\partial_z\delta n\partial_z\overline{\delta n}\text{d}V, \\
    &= -\frac{1}{4}\int\overline{\delta n}\left[K_3\partial^2_z \delta n+2(K_1+K_2)\partial^2_{\omega\bar{\omega}}\delta n+2(K_1-K_2)\partial^2_{\bar{\omega}}\overline{\delta n}\,\right]\text{d}V+\text{c.c.} ,
\end{split}
\end{equation}    
\end{widetext}
where the second line follows from integration by parts. From this we may read off the complex molecular field $h=h_x+\text{i}h_y=-\frac{\delta F}{\delta(\overline{\delta n})}$ as
\begin{equation}
    h=\frac{1}{8}\bar{K}\left[\nabla^2\delta n+\kappa_{\parallel}(\partial^2_z-4\partial^2_{\omega\bar{\omega}})\delta n+8\kappa_{\perp}\partial^2_{\bar{\omega}}\overline{\delta n}\,\right],
    \label{eq:ComplexMolecularField}
\end{equation}
where $\bar{K}=K_3+\frac{K_1+K_2}{2}$ and we have introduced what we term parallel and perpendicular elastic anisotropies $\kappa_{\parallel}=\frac{2K_3-(K_1+K_2)}{2K_3+K_1+K_2}$ and $\kappa_{\perp}=\frac{K_1-K_2}{2K_3+K_1+K_2}$. We name these as such because, in our linearised description, bend corresponds to derivatives parallel to the far-field director while splay and twist correspond to derivatives perpendicular to it. The definitions of these anisotropies are such that $-1\leq \kappa_{\parallel},\kappa_{\perp}\leq 1$. All terms in \eqref{eq:ComplexMolecularField} must belong to the same spin class. Since $\kappa_{\perp}$ appears as a coefficient of $\overline{\delta n}$, this means that, for a given harmonic distortion, the correction due to splay-twist anisotropy has conjugate spin. At monopole order this means splay-twist anisotropy results in a spin $-1$ distortion, and hence results in director perturbations out of the plane defined by the unperturbed monopole. This is illustrated in the top row of Figure \ref{fig:3DFundamentalDistortionsFlows}. At higher orders it necessitates that the transformation of the $\kappa_{\perp}$ term under the action of derivatives is conjugate to that of the isotropic and $\kappa_{\parallel}$ terms.

We now solve $h=0$ perturbatively around the one-elastic-constant solution to first order in the anisotropies. Using $\nabla^2r=\frac{2}{r}$ and writing out the $x$- and $y$-components explicitly we find the modified forms of the fundamental monopole distortions to be
\begin{equation}
    \begin{split}
        \delta\textbf{n}^{(x)}&=\alpha a\left\lbrace\frac{1}{r}\textbf{e}_x+\kappa_{\parallel}\frac{z^2}{r^3}\textbf{e}_x+\kappa_{\perp}\left[\frac{x^2-y^2}{r^3}\textbf{e}_x+\frac{2xy}{r^3}\textbf{e}_y\right]\right\rbrace,\\
        \delta\textbf{n}^{(y)}&=\alpha a\left\lbrace\frac{1}{r}\textbf{e}_y+\kappa_{\parallel}\frac{z^2}{r^3}\textbf{e}_y+\kappa_{\perp}\left[\frac{2xy}{r^3}\textbf{e}_x+\frac{y^2-x^2}{r^3}\textbf{e}_y\right]\right\rbrace.
    \end{split}
    \label{eq:3DFundamentalDirectorModes}
\end{equation}

Under the one-elastic-constant approximation the energy minimising far-field distortions may be written as derivatives of $\frac{1}{r}$, that is $\delta n=\alpha a\mathcal{L}\frac{1}{r}$, where $\mathcal{L}=\mathcal{L}_x+\text{i}\mathcal{L}_y$ is a complex differential operator. The complex spherical harmonics form a natural basis for these distortions and they can be generated by taking $\mathcal{L}$ to be composed of $\partial_z$, which leaves the spin of a distortion unchanged, a spin-lowering operator $\partial_{\bar{\omega}}$ and a spin-raising operator $\partial_{\omega}$ \cite{houston2023active}. Similarly, it follows from \eqref{eq:3DFundamentalDirectorModes} that the leading order form of a generic energy minimising distortion in the presence of elastic anisotropy may be expressed as
\begin{equation}
    \delta n=\alpha a\left\lbrace\mathcal{L}\left[\frac{1}{r}+\frac{1}{2}\kappa_{\parallel}(4\partial^2_{\omega\bar{\omega}}-\partial^2_z)r\right]-4\kappa_{\perp}\bar{\mathcal{L}}\partial^2_{\bar{\omega}}r\right\rbrace.
    \label{eq:FundamentalDirectorModeComplex}
\end{equation}
We highlight that it is the conjugate operator, $\bar{\mathcal{L}}$, that acts on the $\kappa_{\perp}$ correction. As noted before, this stems from $\kappa_{\perp}$ appearing as a coefficient of $\overline{\delta n}$ in \eqref{eq:ComplexMolecularField} and means that under the action of a derivative which, say, lowers the spin of the harmonic distortion, the spin of the splay-twist correction is increased.

The bottom row of Figure \ref{fig:3DFundamentalDistortionsFlows} shows the active flow response associated with the unperturbed monopole and those induced by the director perturbations due to parallel and perpendicular anisotropy. While we derive the latter two in the subsequent section, we can use the former, provided in \eqref{eq:FundamentalActiveFlow3DIsotropic}, to understand their structure on symmetry grounds. The deformations $\delta n\sim1/r$ and $\delta n\sim r$ will induce active flows of the same form, as they are both isotropic. 
Since the director corrections due to parallel and perpendicular anisotropy can be expressed as second-order derivatives of $r$, we therefore expect their active flow responses to be akin to those of the spin $1$ and spin $-1$ quadrupoles respectively \cite{houston2023active}. This correspondence explains both the rotational nature of the flows and the sense of their rotation, which is opposite for the two forms of anisotropy.

Let us briefly summarise our results thus far. Elastic anisotropy may be described by two parameters that range from $-1$ to $1$. One, $\kappa_{\parallel}$, describes the deviation of the bend constant from that of splay and twist. It induces a spin-$1$ correction to the monopole, such that both distortions transform in the same way when differentiated. The other, $\kappa_{\perp}$, captures the difference between the splay and twist constants. It induces a correction to the monopole with spin $-1$ which transforms in a conjugate manner when differentiated, such that for a harmonic distortion with spin $s$, the splay-twist correction has spin $-s$. At leading order the deviation from being harmonic is described by a second-order operator, so that anisotropy naturally induces quadrupole-like perturbations to the monopole, and hence is associated with rotational dynamics.

\subsubsection{Active Response}
As we have seen, a given energy minimising director distortion can be decomposed into a harmonic part, a perturbation due to parallel anisotropy and a perturbation due to perpendicular anisotropy, such that the anisotropy-modified $x$-monopole takes the form
\begin{equation}
    \delta\mathbf{n}^{(x)}=\delta\mathbf{n}^{(x)}_h+\kappa_{\parallel}\delta\mathbf{n}^{(x)}_{\parallel}+\kappa_{\perp}\delta\mathbf{n}^{(x)}_{\perp},
    \label{eq:DirectorDecomposition}
\end{equation}
with the individual components as specified in \eqref{eq:3DFundamentalDirectorModes}. Since we are linearising the active force, the active response can be similarly decomposed according to the component of \eqref{eq:DirectorDecomposition} that induces it, giving the flow as
\begin{equation}
    \tilde{\mathbf{u}}^{(x)}=\tilde{\mathbf{u}}^{(x)}_h+\kappa_{\parallel}\tilde{\mathbf{u}}^{(x)}_{\parallel}+\kappa_{\perp}\tilde{\mathbf{u}}^{(x)}_{\perp},
    \label{eq:FlowDecomposition}
\end{equation}
and similarly for the pressure response. Following the conventions of \cite{houston2023active}, we use tildes to distinguish fundamental responses from those due to particular distortions, while the superscript indicates that these responses are associated with `$x$' director modes. The flow response due to the harmonic part is derived in \cite{houston2023active} and presented in \eqref{eq:FundamentalActiveFlow3DIsotropic}. The remaining active responses are determined by the solutions to the following Stokes equations
\begin{align}
    -\nabla p_{\parallel}+\mu\nabla^2\tilde{\mathbf{u}}_{\parallel}=\partial_x\delta\mathbf{n}_{\parallel}^{(x)}\mathbf{e}_z+\partial_z\delta\mathbf{n}_{\parallel}^{(x)}\mathbf{e}_x,\\
    -\nabla p_{\perp}+\mu\nabla^2\tilde{\mathbf{u}}_{\perp}=\partial_x\delta\mathbf{n}_{\perp}^{(x)}\mathbf{e}_z+\partial_z\delta\mathbf{n}_{\perp}^{(x)}\mathbf{e}_x,
\end{align}
where the right hand sides follow from linearising the appropriate part of the active force.

In solving these Stokes problems we are aided by a fact identified in the previous section: all perturbations to the monopole can be written as derivatives of $r$. Within our linearised framework we can therefore generate all desired active responses from that associated with $\mathbf{n}=\mathbf{e}_z+r\mathbf{e}_x$, for which we find the induced pressure and flow to be
\begin{gather}
    \tilde{P}^{(x)}=-\frac{\zeta xz}{2r} , \\
    \begin{split}
        \tilde{\mathbf{U}}^{(x)} &= \frac{\zeta}{12\mu} \biggl[ \left(2zr-\frac{x^2z}{r}\right)\mathbf{e}_x-\frac{xyz}{r}\mathbf{e}_y \\
        & \qquad +\left(2xr-\frac{xz^2}{r}\right)\mathbf{e}_z \biggr],
    \end{split}
\end{gather}
respectively. The pressure is a scaled version of that for the monopole; this does not hold for the flow, although it contains a contribution of that form and importantly has the same extensional nature.

The desired responses can then be attained via the appropriate derivatives, which may be read off from \eqref{eq:FundamentalDirectorModeComplex}. Explicitly, the fundamental flow response due to a parallel anisotropy perturbation around a monopole in the $x$-direction is
\begin{equation}
    \begin{split}
        \tilde{\mathbf{u}}_{\parallel}^{(x)}&=\alpha a\frac{1}{2}(4\partial^2_{\omega\bar{\omega}}-\partial^2_z)\tilde{\mathbf{U}}^{(x)}\\
    &=\frac{\alpha a\zeta}{12\mu r^5}\Bigl[ 3(x^2+y^2)^2(-z\mathbf{e}_x+x\mathbf{e}_z) \\
    & + 3yz^3(-y\mathbf{e}_x+x\mathbf{e}_y) +r^2z^2(-z\mathbf{e}_x+6x\mathbf{e}_z) \Bigr],
    \end{split}
    \label{eq:FundamentalActiveFlow3DParallel}
\end{equation}
while the corresponding flow due to perpendicular anisotropy is
\begin{equation}
    \begin{split}
        \tilde{\mathbf{u}}_{\perp}^{(x)}&=-\alpha a4\partial^2_{\bar{\omega}}\tilde{\mathbf{U}}^{(x)}\\
        &=\frac{\alpha a\zeta}{12\mu r^5}\left[3xz(x^2+y^2)(x\mathbf{e}_x+y\mathbf{e}_y-z\mathbf{e}_z)\right.\\
    &\left. +4r^2z^2(z\mathbf{e}_x-3x\mathbf{e}_z)-6x(x^2+y^2)^2\mathbf{e}_z\right],
    \end{split}
    \label{eq:FundamentalActiveFlow3DPerp}
\end{equation}
these being the flows that are shown in the bottom row of Figure \ref{fig:3DFundamentalDistortionsFlows}, along with that induced by the harmonic monopole, given in \eqref{eq:FundamentalActiveFlow3DIsotropic}. The corresponding flows associated with a $y$-monopole, $\tilde{\mathbf{u}}_{\parallel}^{(y)}$ and $\tilde{\mathbf{u}}_{\perp}^{(y)}$, are attained by exchanging $x$ and $y$.

\begin{figure*}
    \centering
    \begin{tikzpicture}[scale=1.6,>=stealth]
        \node[anchor=south west,inner sep=0] at (0,0)
{\includegraphics[width=1\linewidth, trim = 10 0 0 0, clip, angle = 0, origin = c]{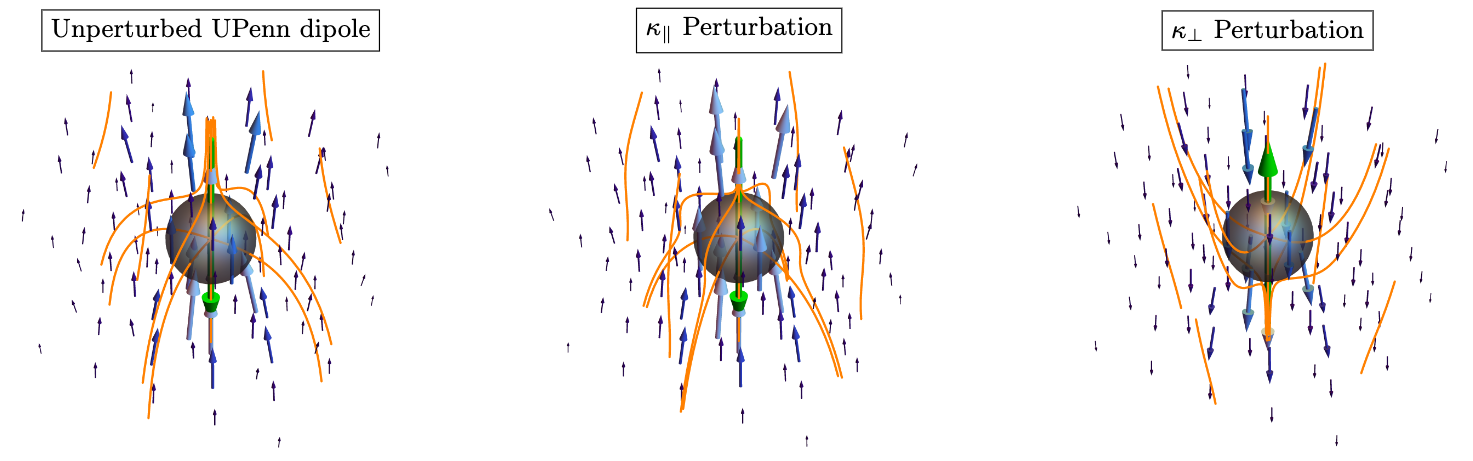}};
    \end{tikzpicture}
    \caption[Caption.]{The active flow response to the UPenn dipole. From left to right: the contribution due to the harmonic distortion, the perturbation due to parallel (bend) anisotropy and that due to perpendicular (splay-twist) anisotropy. In each case the streamlines of the director are shown in orange, the induced active flow is shown with blue arrows and the green arrow indicates the net active force.    }
    \label{fig:UPennFlows}
\end{figure*}

By way of example, we now calculate the leading active flow induced by the UPenn dipole with elastic anisotropy.
The harmonic UPenn dipole is given by
\begin{equation}
    \begin{split}
        \delta\mathbf{n}&=\frac{\alpha a}{2}\left(\partial_x\frac{a}{r}\mathbf{e}_x+\partial_y\frac{a}{r}\mathbf{e}_y\right)=-\frac{\alpha a^2}{2r^3}\left(x\mathbf{e}_x+y\mathbf{e}_y\right),
    \end{split}
\end{equation}
where $\alpha$ is a dimensionless coefficient and $a$ is a characteristic lengthscale of the distortion. In the presence of elastic anisotropy, the leading form of the director is found by replacing the harmonic $1/r$ with the modified forms given in \eqref{eq:3DFundamentalDirectorModes}, while maintaining the same derivative structure. Hence its leading order form is
\begin{equation}
    \begin{split}
        \delta\mathbf{n}&=\frac{\alpha a^2}{2}\left[\partial_x\delta\mathbf{n}^{(x)}+\partial_y\delta\mathbf{n}^{(y)}\right]\\
        &=\frac{\alpha a^2}{2}\left[-\frac{1}{r^3}-\frac{3\kappa_{\parallel}z^2}{r^5}+\frac{\kappa_{\perp}(r^2+3z^2)}{r^5}\right]\left(x\mathbf{e}_x+y\mathbf{e}_y\right).
    \end{split}
\end{equation}
The same set of derivatives and an analogous substitution provides the active flow response. The active flow induced by the director distortion $\delta\mathbf{n}^{(x)}$ is $\tilde{\mathbf{u}}^{(x)}+\kappa_{\parallel}\tilde{\mathbf{u}}^{(x)}_{\parallel}+\kappa_{\perp}\tilde{\mathbf{u}}^{(x)}_{\perp}$, and similarly for the distortion $\delta\mathbf{n}^{(y)}$. Using the fundamental flows given in \eqref{eq:FundamentalActiveFlow3DIsotropic}, \eqref{eq:FundamentalActiveFlow3DParallel} and \eqref{eq:FundamentalActiveFlow3DPerp} gives the UPenn dipole flow as
\begin{widetext}
    \begin{equation}
    \begin{split}
        \mathbf{u}&=\frac{\alpha a^2}{2}\left[\partial_x\left(\tilde{\mathbf{u}}^{(x)}+\kappa_{\parallel}\tilde{\mathbf{u}}^{(x)}_{\parallel}+\kappa_{\perp}\tilde{\mathbf{u}}^{(x)}_{\perp}\right)+\partial_y\left(\tilde{\mathbf{u}}^{(y)}+\kappa_{\parallel}\tilde{\mathbf{u}}^{(y)}_{\parallel}+\kappa_{\perp}\tilde{\mathbf{u}}^{(y)}_{\perp}\right)\right]\\
        &=\frac{\alpha a^2\zeta}{8\mu r^5}\left\lbrace\left[-z(r^2-3z^2)(x\mathbf{e}_x+y\mathbf{e}_y)+(r^4+3z^4)\mathbf{e}_z\right]+\frac{\kappa_{\perp}}{r^2}\left[z^3(r^2-5z^2)(x\mathbf{e}_x+y\mathbf{e}_y)-(2r^6+z^2(r^4+5z^4))\mathbf{e}_z\right]\right.\\
        &\left.+\frac{\kappa_{\parallel}}{r^2}\left[z((r^2-2z^2)^2+z^4)(x\mathbf{e}_x+y\mathbf{e}_y)+(r^6+z^2((x^2+y^2)^2-r^2z^2+4z^4))\mathbf{e}_z\right]
        \right\rbrace.
    \end{split}
\end{equation}
\end{widetext}
The three parts of both the director distortion and the associated flow response are shown in Figure \ref{fig:UPennFlows}. The green arrows in Figure \ref{fig:UPennFlows} indicate the net active forces. The calculation of these, along with net active torques, allows propulsive or rotational dynamics to be identified and will be presented in the next section. While all the distortions induce propulsive flows, that due to perpendicular anisotropy opposes the other two. If $\kappa_{\perp}$ is sufficiently large, to a degree that we quantify in our discussion of active forces, this can reverse the self-propulsion direction of the UPenn dipole.

We conclude this discussion of active responses by noting that, as in \cite{houston2023active}, the formulation of active responses can be usefully complexified via
\begin{widetext}
    \begin{align}
    \tilde{p}&=\frac{\alpha a\zeta z\bar{\omega}}{r^3}\left[1-\frac{\kappa_{\parallel}+\kappa_{\perp}}{2}+(\kappa_{\parallel}-\kappa_{\perp})\frac{3z^2}{2r^2}\right],\\
    \tilde{u}&=\frac{\alpha a\zeta z}{8\mu r}\left\lbrace\frac{\bar{\omega}^2}{r^2}\left[1+\kappa_{\parallel}\frac{z^2}{r^2}+\kappa_{\perp}\frac{\omega\bar{\omega}}{r^2}\right]\mathbf{e}_{\omega}+\left[3\left(1-\frac{z^2}{3r^2}\right)-2\kappa_{\parallel}\left(1-\frac{7z^2}{6r^2}+\frac{z^4}{2r^4}\right)+\kappa_{\perp}\left(1+\frac{2z^2}{3r^2}+\frac{z^4}{r^4}\right)\right]\mathbf{e}_{\bar{\omega}}\right\rbrace\\\nonumber
    &\qquad\qquad+\frac{\alpha a\zeta\bar{\omega}}{4\mu r}\left[\left(1+\frac{z^2}{r^2}\right)+\kappa_{\parallel}\left(1+\frac{z^4}{r^4}\right)-2\kappa_{\perp}\left(1+\frac{z^2}{2r^2}+\frac{z^4}{2r^4}\right)\right]\mathbf{e}_z
    ,
    \end{align}
\end{widetext}
where we have introduced the complex vectors $\mathbf{e}_{\omega}=\mathbf{e}_x+\text{i}\mathbf{e}_y$ and $\mathbf{e}_{\bar{\omega}}=\mathbf{e}_x-\text{i}\mathbf{e}_y$. Here $\tilde{u}$ contains the flow response due to the unperturbed monopole, given in \eqref{eq:FundamentalActiveFlow3DIsotropic}, along with the flows resulting from parallel and perpendicular anisotropy, given in \eqref{eq:FundamentalActiveFlow3DParallel} and \eqref{eq:FundamentalActiveFlow3DPerp}. Similarly, $\tilde{p}$ incorporates the corresponding pressure responses, although these were not given previously. Acting on $\tilde{p}$ and $\tilde{u}$ with a complex differential operator $\mathcal{L}=\mathcal{L}_x+\text{i}\mathcal{L}_y$ produces the responses for a pair of conjugate distortions; the real part giving the response induced by $\delta n=\mathcal{L}_x\delta\mathbf{n}^{(x)}+\mathcal{L}_y\delta\mathbf{n}^{(y)}$, the coefficient of $-\text{i}$ giving that associated with $\delta n=-\mathcal{L}_y\delta\mathbf{n}^{(x)}+\mathcal{L}_x\delta\mathbf{n}^{(y)}$.

\subsubsection{Active forces and torques}
Arguably the most striking effect of elastic anisotropy is, as we have already seen, that it leads to rotational effects associated with the fundamental distortions given in \eqref{eq:3DFundamentalDirectorModes}. For the $\delta\mathbf{n}^{(x)}$ and $\delta\mathbf{n}^{(y)}$monopole the net active torque is given by
\begin{equation}
    \boldsymbol{\tau}^a=8\pi\zeta\alpha aR^2(\kappa_{\parallel}-\kappa_{\perp})\mathbf{e}_y
\end{equation}
and
\begin{equation}
    \boldsymbol{\tau}^a=-8\pi\zeta\alpha aR^2(\kappa_{\parallel}-\kappa_{\perp})\mathbf{e}_x
\end{equation}
respectively. Naturally, the second of these torques is an anticlockwise rotation of the first. That the torque is determined by the difference between the parallel and perpendicular anisotropy coefficients is in accordance with the opposite signs of rotation engendered by each, as highlighted in Figure \ref{fig:3DFundamentalDistortionsFlows}. Since
\begin{equation}
    \kappa_{\parallel}-\kappa_{\perp}=\frac{2(K_3-K_1)}{2K_3+K_1+K_2}
\end{equation}
the induced rotation can be in either sense depending on the form of elastic anisotropy. If we neglect the twist constant, which is typically smaller than the other two, then this term is positive if $K_3>3K_1/2$ and negative otherwise. This degree of bend-dominant anisotropy is within the range that has been observed \cite{zhang2018interplay}.

At higher orders, the matter of which distortions have net active forces or torques due to elastic anisotropy can be answered by appealing to symmetry. The only distortions which can produce a net active force are those with spin $1$ or the achiral spin-$0$ deformation, as discussed in \cite{houston2023active}. Similarly, the only ones associated with net active torques are the chiral spin-$0$ distortion and the four distortions with spin $\pm1$. Since parallel anisotropy leads to a perturbation of the same spin as the harmonic distortion, we expect it to modify all of these forces and torques, without imbuing any new distortions with propulsive or rotational dynamics. The spin of the perpendicular anisotropy perturbation is opposite in sign to the harmonic distortion, meaning that we expect the same as for parallel anisotropy, with the exception of the active forces for spin $1$ dipoles being unaltered.

Performing the calculations, we find that the UPenn dipole generates an active force
\begin{equation}
    \begin{split}
        \mathbf{f}^a&=-\frac{4}{3}\pi\zeta\alpha a^2\left(1+\frac{3}{5}\kappa_{\parallel}-\frac{8}{5}\kappa_{\perp}\right)\mathbf{e}_z\\
        &=\frac{4\pi\zeta\alpha a^2}{15\bar{K}}\left(3K_1-5K_2-8K_3\right)\mathbf{e}_z,
    \end{split}
\end{equation}
while the spin-$1$ dipoles along $\mathbf{e}_i$, $i=x,y$ produce
\begin{equation}
    \mathbf{f}^a=-\frac{4}{3}\pi\zeta\alpha a^2\left(1-\frac{1}{5}\kappa_{\parallel}\right)\mathbf{e}_i.
\end{equation}
The propulsion direction of the spin-$1$ dipoles is unchanged by elastic anisotropy. However, the active force acting on the UPenn dipole, which is distinguished by being modified by perpendicular anisotropy, can be reversed. For equal elastic constants it is along $-\mathbf{e}_z$, but, again neglecting the twist constant, changes sign if $K_1>8K_3/3$. This threshold is close to observed values of anisotropy of $K_1/K_3\approx2$ \cite{zhang2018interplay}. In the limit where the elastic energy is dominated by splay ($K_2=K_3=0$) the net active force becomes $\frac{8}{5}\pi\zeta\alpha a^2\mathbf{e}_z$. This would act to increase the distance between a colloid and its satellite defect and could in principle lead to their separation.

\begin{table*}
\begin{center}
\begin{tabular}{| c | Sc | Sc |} 
 \hline
 \textbf{Spin} & \textbf{Distortion} & \textbf{Net Active Torque} \\ [0.5ex] 
 \hline
 -1 & $\frac{a^2}{4}\left[\left(\partial^2_x-\partial^2_y\right)\delta\mathbf{n}^{(x)}+2\partial^2_{xy}\delta\mathbf{n}^{(y)}\right]$ & $-\frac{4}{5}\pi\zeta\alpha a^3\left(1+\frac{5}{7}\kappa_{\parallel}-\frac{8}{7}\kappa_{\perp}\right)\mathbf{e}_y$\\
 \hline
 -1 & $\frac{a^2}{4}\left[-2\partial^2_{xy}\delta\mathbf{n}^{(x)}+(\partial^2_x-\partial^2_y)\delta\mathbf{n}^{(y)}\right]$ & $-\frac{4}{5}\pi\zeta\alpha a^3\left(1+\frac{5}{7}\kappa_{\parallel}-\frac{8}{7}\kappa_{\perp}\right)\mathbf{e}_x$ \\
 \hline
 0 & $\frac{a^2}{2}(-\partial^2_{yz}\delta\mathbf{n}^{(x)}+\partial^2_{xz}\delta\mathbf{n}^{(y)})$ & $\frac{4}{5}\pi\zeta\alpha a^3\left(1+\frac{1}{7}\kappa_{\parallel}+\frac{8}{7}\kappa_{\perp}\right)\mathbf{e}_z$ \\
 \hline
 1 & $a^2\partial^2_z\delta\mathbf{n}^{(x)}$ & $\frac{8}{5}\pi\zeta\alpha a^3\left(1-\frac{5}{7}\kappa_{\parallel}+\frac{4}{7}\kappa_{\perp}\right)\mathbf{e}_y$ \\
 \hline
 1 & $a^2\partial^2_z\delta\mathbf{n}^{(x)}$ & $-\frac{8}{5}\pi\zeta\alpha a^3\left(1-\frac{5}{7}\kappa_{\parallel}+\frac{4}{7}\kappa_{\perp}\right)\mathbf{e}_x$ \\ [1ex] 
 \hline
\end{tabular}
\end{center}
\caption{Active torques associated with quadrupoles and their anisotropic perturbations. The distortions are presented as derivatives of the fundamental director distortions given in \eqref{eq:3DFundamentalDirectorModes}. As befits their spin class, the successive distortions, and hence torques, with spin $-1$ and $1$ are related by clockwise and anticlockwise rotations respectively.}
\label{Table:Quadrupoles}
\end{table*}

Again, the set of elastically perturbed quadrupoles that generate net active torques is the same five that do so in the unperturbed, elastically isotropic, case. The net active torques are summarised in Table I. In all cases the sign of the torque can be reversed by elastic anisotropy. For quadrupoles with spin $-1$, $0$ and $1$ this reversal is maximised when the distortion energy is dominated by the splay, twist and bend elastic constant respectively.

\subsection{Two dimensions}
As in our investigation of three-dimensional distortions, we establish a dimensionless parameter that measures the difference in the splay and bend elastic constants, determine the leading director perturbations due to this parameter being non-zero and then calculate the associated active response. We again find anomalous active torques associated with the monopole, the underlying logic of which is the same as illustrated in Figure \ref{fig:AnisotropicEnergy} for the three-dimensional case. With regards to propulsive motion, a given sign of elastic anisotropy enhances the net force associated with one dipole and diminishes that associated with the other, as can be explained by the dominantly splay or bend character of the dipoles.

\subsubsection{Morphology}
In two dimensions the far-field director has the form $\mathbf{n}=\mathbf{e}_y+\delta n\mathbf{e}_x$. Consequently the splay and bend are given, to linear order in $\delta n$, by $\partial_x\delta n$ and $\partial_y \delta n$ respectively and the Frank free energy may be written as
\begin{equation}
    F=\frac{\Bar{K}}{2}\int\left[(1-\kappa)(\partial_x\delta n)^2+(1+\kappa)(\partial_y\delta n)^2\right],
    \label{AnisotropicFrank2D}
\end{equation}
where $\Bar{K}=(K_1+K_3)/2$ is the mean elastic constant and $\kappa=(K_3-K_1)/(K_1+K_3)$ is the anisotropy parameter. When $\kappa=0$ \eqref{AnisotropicFrank2D} is a Dirichlet energy, with the corresponding Euler-Lagrange equation given by Laplace's equation -- for which the general solution is the sum of a holomorphic and antiholomorphic function and the `fundamental solution' is the nematic monopole $\delta n\sim\ln r$. It is clear that the effect of elastic anisotropy is to scale the two coordinates such that the nematic monopole is now given by
\begin{equation}
    \delta n=\alpha\frac{\ln\left(\sqrt{\frac{x^2}{1-\kappa}+\frac{y^2}{1+\kappa}}/R\right)}{\ln(a/R)},
    \label{eq:2DFarFieldExact}
\end{equation}
where $\alpha$ is again a dimensionless coefficient. This scaling of $x$ and $y$ is suggestive of a connection to quasiconformal maps \cite{ahlfors2006lectures}.

\subsubsection{Active Response}
\label{subsubsec:2DActiveResponse}
As in the three-dimensional case, the active response can be decomposed into that due to the elastically isotropic distortion and the part due to the elastic anisotropy perturbation. The flows that we derive and present here do not represent the full active response due to a given distortion, but rather the leading correction to the response under a one-elastic-constant approximation, which was determined in \cite{houston2023active}.

Expanding \eqref{eq:2DFarFieldExact} to leading order in $\kappa$ gives
\begin{equation}
    \begin{split}
        \delta n&=\frac{\alpha}{\ln(a/R)}\left[\ln(r/R)+\kappa\frac{x^2-y^2}{2r^2}\right]\\
    &=\alpha\left[\frac{\ln(r/R)}{\ln(a/R)}+\kappa\frac{1}{4}\left(\partial^2_x-\partial^2_y\right)\frac{r^2\ln(r/R)}{\ln(a/R)}\right].
    \label{eq:DisortionLeadingOrder2D}
    \end{split}
\end{equation}
We therefore need only solve for the active flow induced by $\delta n=r^2\ln(r/R)/\ln(a/R)$. With this choice of director we may write the two-dimensional analogue of the Stokes equation with active forcing, given in \eqref{eq:StokesActiveForce}, as
\begin{equation}
    \begin{split}
        2\partial_{\bar{z}}(-\tilde{P}+i\mu\tilde{\Omega})=\zeta\partial_{\bar{z}}\left[\frac{i\bar{z}^2}{\ln(a/R)}\left(\frac{1}{4}+\ln(r/R)\right)\right],
    \end{split}
\end{equation}
where here $z=x+iy$. Reading off the real and imaginary parts provides the pressure and vorticity as
\begin{align}
    \tilde{P} &= -\frac{\zeta xy}{\ln(a/R)}\left(\frac{1}{4}+\ln(r/R)\right), \\
    \tilde{\Omega} &= \frac{\zeta(x^2-y^2)}{2\mu\ln(a/R)}\left(\frac{1}{4}+\ln(r/R)\right).
\end{align}
Since the flow is incompressible we have $\tilde{\Omega}=-4\partial_z\partial_{\bar{z}}\tilde{\Psi}=-2i\partial_z\tilde{U}$, with $\tilde{\Psi}$ the streamfunction, from which we find the flow to be
\begin{equation}
    \begin{split}
        \tilde{\mathbf{U}}&=\frac{\zeta}{72\mu\ln(a/R)} \Bigl[ -2r^2(y\mathbf{e}_x+x\mathbf{e}_y) \\
        & -xy(x\mathbf{e}_x+y\mathbf{e}_y)+12\ln(r/R)(y^3\mathbf{e}_x+x^3\mathbf{e}_y) \Bigr].
    \end{split}
\end{equation}
Applying the appropriate derivatives and prefactors gives the leading contribution to the active response due to elastic anisotropy as
\begin{gather}
    \tilde{p}=\frac{\alpha\zeta\kappa xy(x^2-y^2)}{2r^4\ln(a/R)},
    \label{eq:AnisotropicMonopolePressure2D} \\
    \begin{split}
        \tilde{\mathbf{u}}&=\frac{\alpha\zeta\kappa}{48\mu\ln(a/R)} \biggl[ (7+12\ln(r/R))(-y\mathbf{e}_x+x\mathbf{e}_y) \\
        & \quad \qquad +\frac{8}{r^4}(yx^4\mathbf{e}_x-xy^4\mathbf{e}_y) \biggr].
    \end{split}
    \label{eq:AnisotropicMonopoleFlow2D}
\end{gather}

\begin{figure}
    \centering
    \begin{tikzpicture}[scale=1.6,>=stealth]
        \node[anchor=south west,inner sep=0] at (0,0)
{\includegraphics[width=0.95\linewidth, trim = 10 0 0 0, clip, angle = 0, origin = c]{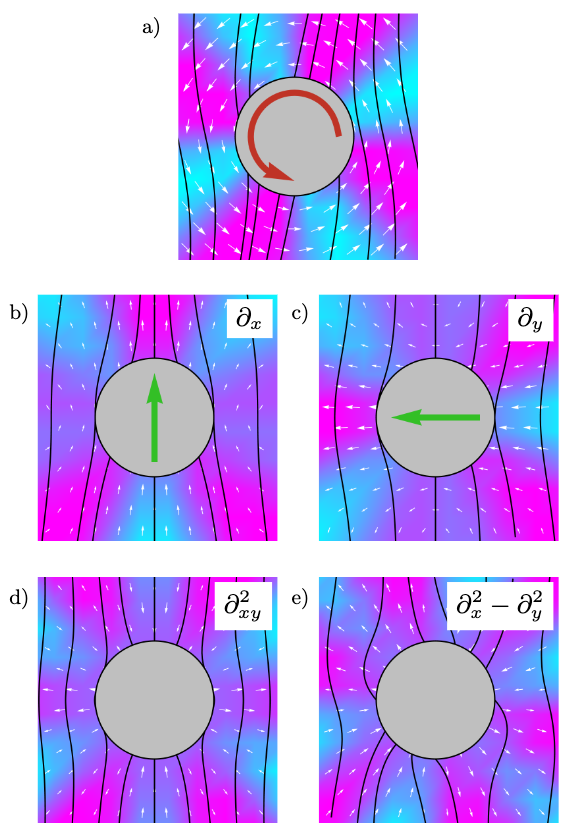}};
        
    \end{tikzpicture}
    \caption[Distortions up to quadrupole order in two-dimensional active nematics with elastic anisotropy.]{Leading corrections due to elastic anisotropy for distortions up to quadrupole order in two-dimensional active nematics. The integral curves of the director (black lines) and the active flow (white arrows) are superposed on the pressure field. The panels are labelled according to the appropriate derivative of the fundamental distortion. The red and green arrows indicate the self-rotation or -propulsion that would result in an extensile system. For both the director and the associated active response we show only the leading contribution due to elastic anisotropy, neglecting the harmonic part. Therefore the director stems from the coefficient of $\kappa$ in \eqref{eq:DisortionLeadingOrder2D}, while the active response is derived from \eqref{eq:AnisotropicMonopolePressure2D} and \eqref{eq:AnisotropicMonopoleFlow2D}.}
    \label{fig:ActiveFlows2D}
\end{figure}

The pressure and flow induced by anisotropy at monopole order, given in \eqref{eq:AnisotropicMonopolePressure2D} and \eqref{eq:AnisotropicMonopoleFlow2D}, are shown in Figure \ref{fig:ActiveFlows2D}a), along with the perturbation to the director from \eqref{eq:DisortionLeadingOrder2D}. Here the effect of anisotropy in two dimensions is analogous to that of bend anisotropy in three dimensions, captured by $\kappa_{\parallel}$. The director perturbation has the symmetry of the chiral quadrupole, the addition of which reduces the bend in the monopole distortion and induces a rotational active flow. The deformations and corresponding active responses at higher orders are shown in the remaining panels of Figure \ref{fig:ActiveFlows2D}, labelled by the requisite derivatives of \eqref{eq:DisortionLeadingOrder2D}, \eqref{eq:AnisotropicMonopolePressure2D} and \eqref{eq:AnisotropicMonopoleFlow2D}. 

At dipole order the anisotropic corrections induce self-propulsive active flows, as shown in Figure \ref{fig:ActiveFlows2D}b) and c) for the vertical dipole (aligned parallel to the far-field director) and horizontal dipole (aligned perpendicular to the far-field director) respectively. Comparison with the active response for elastically isotropic dipoles \cite{houston2023active,houston2023colloids}, shows that for positive elastic anisotropy the self-propulsion of the vertical dipole is increased but that of the horizontal dipole is diminished. These opposing effects have a simple interpretation: the dipoles along and perpendicular to the far field director are comprised of mainly splay and bend distortions respectively, the former is enhanced for $\kappa>0$, the latter for $\kappa<0$. These dipoles are the far-field distortions around oppositely-charged defects, and so the observation that changing their orientation changes the proportion of splay and bend anticipates the elastic torque that acts on defect pairs due to elastic anisotropy, which we shall consider in \ref{subsec:DefectPairs}.

In a deviation from the three-dimensional case, the anisotropic correction to the chiral quadrupole, shown in Figure \ref{fig:ActiveFlows2D}e) does not induce a rotational flow. In fact, both it and the perturbation to the achiral quadrupole, shown in Figure \ref{fig:ActiveFlows2D}d), have a purely radial active flow response, as can be readily found by applying the appropriate derivatives to \eqref{eq:AnisotropicMonopoleFlow2D}.

\subsubsection{Active forces and torques}
As in the three-dimensional case, the greatest departure from isotropic systems is that the perturbation to the monopole distortion induces a net active torque, as evidenced by the rotational flow in Figure \ref{fig:ActiveFlows2D}a). Integrating the active stresses over a circle of radius $r$ gives this net active torque as
\begin{equation}
    \tau=\frac{\zeta\pi\alpha}{2\ln(a/R)}\left[\kappa r^2c_0-4a^2c_{x^2-y^2}\right],
\end{equation}
with $c_0$ and $c_{x^2-y^2}$ the coefficients of $\delta n$ and $(\partial_x^2-\partial_y^2)\delta n$ respectively. These two contributions to the active torque may be viewed as defining a lengthscale, $\sim a/\sqrt{|\kappa|}$, within which the torque is the standard quadrupolar contribution \cite{houston2023active,houston2023colloids}, and beyond which the aniosotropic monopole term dominates. As the anisotropic contribution is connected to the monopole, it does not arise for the idealised case of a disc with tilted anchoring, but is generically present for example for a chiral cog \cite{houston2023colloids,ray2023rectified}, for which it may enhance or impede active rotation.

At dipole order we find an anisotropy-induced correction to the net active force
\begin{equation}
    \mathbf{f}^a=\frac{\zeta\pi\alpha a}{\log(a/R)}\left[\left(1+\frac{\kappa}{2}\right)c_x\mathbf{e}_y+\left(1-\frac{\kappa}{2}\right)c_y\mathbf{e}_x\right],
    \label{eq:ActiveForceDipole2D}
\end{equation}
where $c_i$ denotes the coefficient of $a\partial_i\delta n$. Positive elastic anisotropy enhances the force along the director but reduces the perpendicular component. This mirrors the propulsive flows shown in Figure \ref{fig:ActiveFlows2D}b) and c), and has the same explanation. Note that since $-1\leq\kappa\leq 1$ the effect of elastic anisotropy can never reverse the propulsion direction, the most it can do is halve the speed. Correspondingly, it can increase the speed by at most a factor of $3/2$. In elastically isotropic systems a dipole aligned at an angle $\chi$ to the far field self-propels at an angle $-\chi$. With elastic anisotropy this no longer holds generically (it is only true for pure dipoles, i.e. when $\chi$ is a multiple of $\pi/2$).

\begin{figure}
    \centering
    \begin{tikzpicture}[scale=1.6,>=stealth]
        \node[anchor=south west,inner sep=0] at (0,0)
{\includegraphics[width=1\linewidth, trim = 0 0 0 0, clip, angle = 0, origin = c]{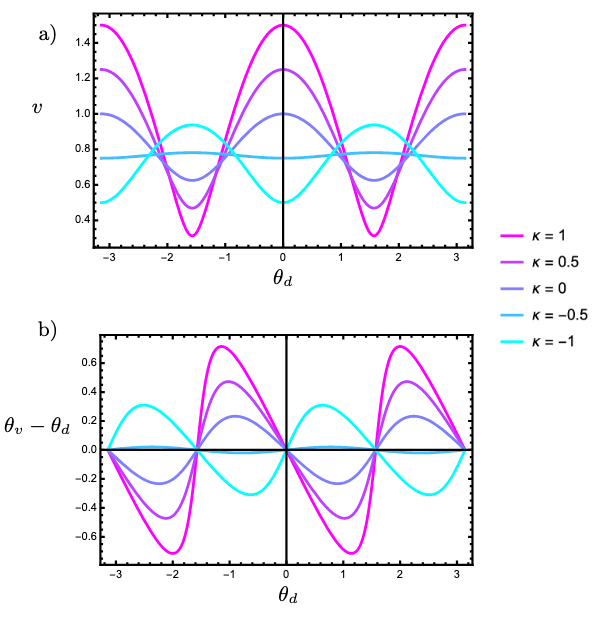}};

    \end{tikzpicture}
    \caption[]{Active propulsion velocity as a function of dipole distortion orientation for different values of elastic anisotropy $\kappa$. a) The propulsive speed, $v$, given by the magnitude of $\mathbf{v}$ in \eqref{eq:ActiveVelocityDipole2D}, normalised by the prefactor $\zeta\alpha c/(6\mu_{\parallel}\log(a/R))$. The orientation of the dipole distortion, $\theta_d$, is measured anti-clockwise from the far-field direction. b) The misalignment, in radians, between the dipole orientation and that of the active velocity, $\theta_v$.
    }
    \label{fig:ActivePropulsion2D}
\end{figure}

To explore further the impact of elastic anisotropy on active propulsion we consider a generic dipole of the form $c_x=c\cos\theta_d$, $c_y=c\sin\theta_d$, from which the active force follows via \eqref{eq:ActiveForceDipole2D}. This force is related to a propulsive velocity by a mobility matrix \cite{kim2013microhydrodynamics}, which in the case of passive nematics has been shown by experiments \cite{loudet2004stokes} and simulations \cite{ruhwandl1996friction,stark2001stokes} to be well-approximated by a diagonal form, with distinct viscosities $\mu_{\parallel}$ and $\mu_{\perp}$, relating to motion parallel and perpendicular to the far-field director respectively. Incorporating this viscous anisotropy yields the velocity
\begin{equation}
    \mathbf{v}=-\frac{\zeta\alpha c}{6\log(a/R)}\left[\frac{\left(1+\frac{\kappa}{2}\right)\cos\theta_d}{\mu_{\parallel}}\mathbf{e}_y+\frac{\left(1-\frac{\kappa}{2}\right)\sin\theta_d}{\mu_{\perp}}\mathbf{e}_x\right].
    \label{eq:ActiveVelocityDipole2D}
\end{equation}
From this expression for $\mathbf{v}$ we may extract the magnitude, $v$, and direction, $\theta_v$, of active propulsion, which are shown in Figure \ref{fig:ActivePropulsion2D} for a typical viscosity ratio of $\mu_{\perp}/\mu_{\parallel}=1.6$ \cite{loudet2004stokes,ruhwandl1996friction,stark2001stokes}. Both the elastic and viscous anisotropy cause variations in the speed with respect to $\theta_d$, shown in Figure \ref{fig:ActivePropulsion2D}a), and lead to the dipole distortion and active propulsion generically being misaligned, as captured in the behaviour of $\theta_v-\theta_d$ in Figure \ref{fig:ActivePropulsion2D}b). As mentioned above, this misalignment vanishes only when $\theta_d$ is a multiple of $\pi/2$. Viscous anisotropy with $\mu_{\perp}>\mu_{\parallel}$ plays a similar role to elastic anisotropy with $K_3>K_1$, namely they both promote motion along the far-field direction, the former by reducing the resistance, the latter by increasing the force. Hence the variation with respect to $\theta_d$ in Figure \ref{fig:ActivePropulsion2D} is more pronounced for $\kappa>0$ than for $\kappa<0$. Provided $\mu_{\perp}/\mu_{\parallel}\leq3$, it is possible for the viscous anisotropy to be compensated by negative elastic anisotropy, that is an energetic preference for bend, with $\kappa=2(1-\mu_{\perp}/\mu_{\parallel})/(1+\mu_{\perp}/\mu_{\parallel})$.

Despite having the appropriate symmetry, the perturbation to the chiral quadrupole does not make a contribution to the net active torque. This is in accordance with the radial flow shown in Figure \ref{fig:ActiveFlows2D}e).

\section{Two-dimensional defects}
We now turn our attention from far-field distortions to topological defects, focusing on two dimensions. As in Section \ref{sec:Far-field}, we shall proceed by first establishing the effect of elastic anisotropy on the morphology, before then determining the consequences this has for active dynamics. Lastly, we consider how elastic anisotropy might alter the behaviour of multi-defect states.

\subsection{Defect morphology}
The morphology of defects in elastically anisotropic two-dimensional nematics has been determined by Dzyaloshinkii \cite{dzyaloshinskii1970theory}, as summarised by Landau and Lifshitz \cite{landau1995course}. Further considerations have been carried out in the context of passive materials \cite{ranganath1983energetics,chandrasekhar1986structure,hudson1989frank}, in part motivated by the high elastic anisotropy that arises close to the nematic-smectic transition \cite{chandrasekhar1986structure}.
We begin by formulating the Frank free energy in such a way as to highlight a certain symmetry in the defect structure with regard to changes in the elastic constants. Although this symmetry has been noted before \cite{hudson1989frank}, our presentation here helps elucidate its origin. This symmetry will prove important in the behaviour of active nematics and further motivates our choice of elastic anisotropy parameter.

\begin{figure}
    \centering
        \begin{tikzpicture}
            \node[anchor=south west,inner sep=0] at (0,0)
{\includegraphics[width=1\linewidth, trim = 0 0 0 0, clip]{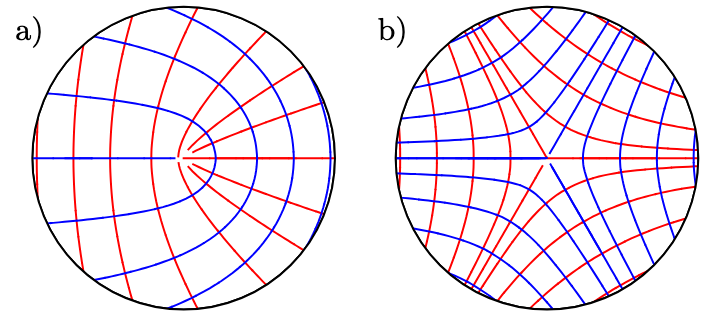}};
            
    \end{tikzpicture}
    \caption{The director fields for equal and opposite values of the elastic anisotropy parameter $\kappa$ are related through othogonality. The integral curves of the director field are shown for $\kappa=-0.5$ (blue) and $\kappa=0.5$ (red) for a) a $+1/2$ defect and b) a $-1/2$ defect.}
    \label{OrthogonalitySymmetry}
\end{figure}

Isolating the energy contribution due to elastic anisotropy, we write the Frank free energy as
\begin{equation}
    F=\frac{\Bar{K}}{2}\int\text{d}A|\nabla\mathbf{n}|^2+\kappa\left[|(\mathbf{n}\cdot\nabla)\mathbf{n}|^2-(\nabla\cdot\mathbf{n})^2\right].
    \label{FrankElasticAnisotropy}
\end{equation}
Note that this is the general form of the energy given in \eqref{AnisotropicFrank2D}, where, owing to our far-field description, bend and splay deformations appeared as $\partial_y\delta n$ and $\partial_x\delta n$ respectively. Considering the field $\mathbf{n}_{\perp}$ that is everywhere orthogonal to $\mathbf{n}$, it is clear that the deformations of the two are linked and in flat two-dimensional space one finds that $(\mathbf{n}_{\perp}\cdot\nabla)\mathbf{n}_{\perp}=-(\nabla\cdot\mathbf{n})\mathbf{n}$, that is the bend of one is proportional to the splay of the other. Consequently we may re-express the Frank free energy as
\begin{equation}
    F=\frac{\Bar{K}}{2}\int\text{d}A|\nabla\mathbf{n}|^2+\kappa\left[|(\mathbf{n}\cdot\nabla)\mathbf{n}|^2-|(\mathbf{n}_{\perp}\cdot\nabla)\mathbf{n}_{\perp}|^2\right]
\end{equation}
and, since $|\nabla\mathbf{n}|^2=|\nabla\mathbf{n}_{\perp}|^2$, see that reversing the sign of $\kappa$ is equivalent to replacing $\mathbf{n}$ with $\mathbf{n}_{\perp}$. The energy-minimising defect morphologies for opposite signs of $\kappa$ are therefore related through orthogonality, as illustrated in Figure \ref{OrthogonalitySymmetry}. This symmetry comes with the caveat that it reverses the defect orientation \cite{vromans2016orientational,tang2017orientation}, which for $+1/2$ defects is given by
\begin{equation}
    \mathbf{p}=\frac{\nabla\cdot(\mathbf{n}\mathbf{n})}{|\nabla\cdot(\mathbf{n}\mathbf{n})|},
    \label{DefectOrientation}
\end{equation}
but since the defect energy is invariant to global rotations we can simply rotate the whole texture by $\pi$ to restore the original orientation.

This symmetry provides an additional motivation for our choice of elastic anisotropy parameter. Determining how it impacts the defect structure is facilitated by specifying a coordinate system and there are three natural candidates: Cartesians using the defect orientation to determine the $x$-axis, polar coordinates centred on the defect or coordinates based on the orthogonal lines of $\mathbf{n}$ and $\mathbf{n}_{\perp}$. Intriguingly no one of these is optimal for describing all aspects of the active dynamics of nematic defects and we shall have cause to use each of them. Presently though, polar coordinates, $(r,\theta)$, are the most useful and so we describe the director through the angle $\psi$ that it makes with the radial line from the defect such that
\begin{equation}\label{DirectorPolar}
    n_r=\cos\psi,\qquad n_{\theta}=\sin\psi.
\end{equation}
For an isolated defect in an infinite system $\psi$ is a function of $\theta$ only, there being no lengthscale to combine with $r$ to form a dimensionless quantity. In the absence of elastic anisotropy $\psi=(s-1)\theta$, making it clear that index $s$ defects have $2|s-1|$ radial lines and possess the corresponding $2|s-1|$-fold rotational symmetry. Imposing elastic anisotropy should not change any of the spatial symmetries of a defect. It is shown in \cite{dzyaloshinskii1970theory,landau1995course} that using \eqref{DirectorPolar} in \eqref{FrankElasticAnisotropy} yields an Euler-Lagrange equation
\begin{equation}
    (1-\kappa\cos2\psi)\psi''=\kappa\sin2\psi(1-\psi'^2),
    \label{eq:ELDefectDirector}
\end{equation}
with a first integral
\begin{equation}
    (1-\kappa\cos2\psi)(\psi'^2-1)=\text{constant},
    \label{eq:FirstIntegral}
\end{equation}
so that $\psi$ may be determined as
\begin{align}
    \theta&=q\int_0^{\psi}\sqrt{\frac{1-\kappa\cos2x}{1-\kappa q^2\cos2x}}\text{d}x,
\end{align}
with $q$ a constant determined by the condition
\begin{align}
    (s-1)q\int_0^{\pi}\sqrt{\frac{1-\kappa\cos2\psi}{1-\kappa q^2\cos2\psi}}\text{d}\psi=\pi.
\end{align}
Note that this solution is valid for $s\neq1$. For integer defects any discrepancy in elastic constants will cause the energy minimiser to be either an aster, with pure splay, or a vortex, with pure bend.

\subsection{Defect dynamics}
\begin{figure*}
  \centering
    \begin{tikzpicture}
\node[anchor=south west,inner sep=0] at (0,0)
{\includegraphics[width=1\linewidth, trim = 0 0 0 0, clip]{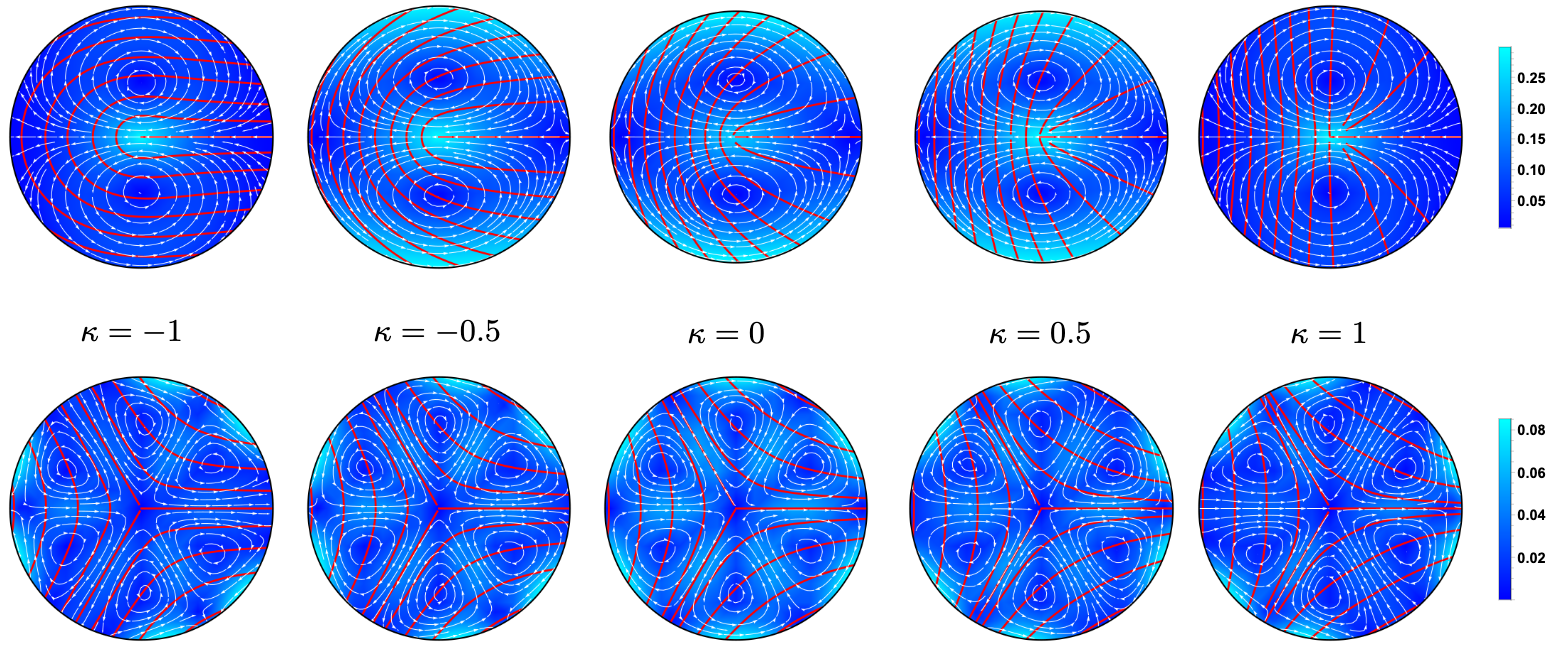}};

\end{tikzpicture}
    \caption{The active flows around nematic point defects. Top row: the flows to second order around $+1/2$ defects. Bottom row: the flows to first order around $-1/2$ defects. The flow streamlines are shown in white, the background colour is lighter when the flow velocity is higher, in accordance with the scale bars, and the red lines indicate the perturbative approximation to the director profile. Note that the scale bars are different for the two sets of defects, as the flow velocity is typically higher around $+1/2$ defects.}
    \label{fig:ActiveFlowsPerturbative}
\end{figure*}
We are now in a position to calculate perturbatively the active backflow of nematic disclinations with unequal elastic constants. Similar calculations have been performed previously for equal elastic constants \cite{giomi2014defect} by convolving the active force with the two-dimensional Oseen tensor \cite{di2008hydrodynamic}. The the same approach is employed here, and as in \cite{giomi2014defect} we consider the defect to be at the centre of a circular domain of radius $R$, with this length characterising the distance to the nearest defect. We relegate the details of the calculation to Appendix A.1 and here simply present the results.

The active flows with the leading contribution due to elastic anisotropy are, for $+1/2$ and $-1/2$ defects respectively,
\begin{widetext}
    \begin{align}
    \begin{split}
        \mathbf{u}_+(\mathbf{r})&=\frac{\zeta}{12\mu}\left[\left(3(r-R)-r\cos2\theta\right)\hat{\mathbf{e}}_x-r\sin2\theta\hat{\mathbf{e}}_y\right]+\frac{\kappa^2\zeta}{512R\mu}\left\lbrace\left[9(R^2+6Rr-7r^2)+8r(3r-R)\cos2\theta \right.\right. \label{eq:ActiveFlowPlusHalf}\\
    & \left.\left.  +2Rr\cos4\theta\right]\hat{\mathbf{e}}_x+\left[r(39r-28R+4R\cos2\theta)\sin2\theta\right]\hat{\mathbf{e}}_y\right\rbrace,
    \end{split}
    \\
    \begin{split}
        \mathbf{u}_-(\mathbf{r})&=\frac{\zeta r}{240R\mu}\left[(5(4R-3r)\cos2\theta+4R\cos4\theta)\hat{\mathbf{e}}_x + (-5(4R-3r)\sin2\theta+4R\sin4\theta)\hat{\mathbf{e}}_y
    \right] \label{eq:ActiveFlowMinusHalf}\\
    &+\frac{\kappa\zeta r}{1728R^4\mu}\left\lbrace\left[2(5R^4-3r^4)\cos5\theta+5R^4\cos7\theta\right]\hat{\mathbf{e}}_x+\left[2(3r^4-5R^4)\sin5\theta+5R^4\sin7\theta\right]\hat{\mathbf{e}}_y\right\rbrace.
    \end{split}
\end{align}
\end{widetext}
These flow fields are illustrated in Figure \ref{fig:ActiveFlowsPerturbative}. The bottom row shows that elastic anisotropy reduces the six-fold symmetry of the flow induced by the $-1/2$ defect to a three-fold symmetry, while the top row shows that the flow of the $+1/2$ defect, and in particular the flow at the origin, corresponding to the defect motility, is much less variable. This accords with the absence of a term proportional to $\kappa$ in the flow induced by the $+1/2$ defect \eqref{eq:ActiveFlowPlusHalf}. Furthermore, the flows for the $+1/2$ defects with equal and opposite values of elastic anisotropy are identical. While this is natural in Figure \ref{fig:ActiveFlowsPerturbative} since the perturbative flow in \eqref{eq:ActiveFlowPlusHalf} only contains $\kappa$ at quadratic order, this is in fact an exact identity necessitated by the symmetry of the $+1/2$ defect. We now present two arguments to explain these observations.

The first is a geometric argument for the flows to first order in $\kappa$. At this level of perturbation the change in the director field is bound by two constraints: it must keep the director unit length and so be everywhere orthogonal to the harmonic director field, and it must maintain the rotational symmetry of the defect. Hence the director field around an index $s$ defect is
\begin{align}
    \mathbf{n}&=\cos s\theta\hat{\mathbf{e}}_x+\sin s\theta\hat{\mathbf{e}}_y\\ \nonumber
    &+\beta\kappa\sin\left(2|s-1|\theta\right)(-\sin s\theta\hat{\mathbf{e}}_x+\cos s\theta\hat{\mathbf{e}}_y)+O(\kappa^2),
\end{align}
where $\beta$ is a constant not determined by our geometric argument. This results in an active force
\begin{widetext}
    \begin{align}
    \mathbf{f}&=\frac{s}{r}\left[\cos((2s-1)\theta)\mathbf{e}_x+\sin((2s-1)\theta)\mathbf{e}_y\right]+\frac{2\beta\kappa}{r}\left\lbrace |s-1|\cos(2|s-1|\theta)\left[\cos((2s-1)\theta)\mathbf{e}_x+\sin((2s-1)\theta)\mathbf{e}_y\right]\right.\\ \nonumber
    &\left. \qquad\qquad\qquad+s\sin(2|s-1|\theta)\left[-\sin((2s-1)\theta)\mathbf{e}_x+\cos((2s-1)\theta)\mathbf{e}_y\right]
    \right\rbrace+O(\kappa^2).
\end{align}
\end{widetext}
From this we find that for $+1/2$ defects the first order contribution to the active force is purely radial. This can be compensated by a radial pressure gradient without setting up any flows. Since the active force $\sim\frac{1}{r}$ the pressure is logarithmically divergent at the defect itself. Therefore $\kappa$ only appears at second order in the active flow \eqref{eq:ActiveFlowPlusHalf}, as demonstrated through the remarkable consistency of the flows for different defect morphologies shown in the top row of Figure \ref{fig:ActiveFlowsPerturbative}. Considering the $-1/2$ defect in the same fashion we find that the first order correction to the active force is composed of a purely radial part and an index $-5$ component, the former again contributing only to a pressure gradient. When solving for the active flow with a no-slip condition on a disc, an active force with index $m$ will produce two flow components, one with index $m$, the other with index $2-m$. Hence the first order contribution to the active flow around a $-1/2$ defect is a combination of an index $-5$ and an index $+7$ flow. Both of these flows consist of twelve lobes, since winding with a non-zero index $s$ is associated with $2|s-1|$ separatrices, as opposed to the six of the elastically isotropic case. This results in the six-fold rotational symmetry of the flow magnitude being broken to three-fold symmetry, matching that of the defect itself. As can be seen in the bottom row of Figure \ref{fig:ActiveFlowsPerturbative}, when $\kappa<0$ and splay deformations are enhanced the flow lines are stretched along the local director direction, as expected since splay produces an active force along the director. When $\kappa>0$ the converse is true with the prevalence of bend causing the flow field to be stretched perpendicular to the director field. While the rotational symmetry of the flow magnitude has been reduced by the introduction of elastic anisotropy, the flow field itself has three-fold symmetry in all cases. As the effect of elastic anisotropy is spatially homogeneous it is incapable of removing this symmetry and thus can never induce active motility in the $-1/2$ defect.

The second argument is one that relates the active flows for different values of elastic anisotropy. As we have already observed, the director fields which minimise the Frank free energy for oppositely signed values of $\kappa$ are related through orthogonality, as is shown in Figure \ref{OrthogonalitySymmetry}. Under this orthogonality condition the sign of the deviatoric part of the active stress tensor, proportional to $\mathbf{n}\mathbf{n}-\mathbf{n}_{\perp}\mathbf{n}_{\perp}$, changes, reversing both the active stresses, and hence active flows, as well as the defect orientation given for $+1/2$ defects by \eqref{DefectOrientation}. The active flows of defects with equal and opposite values of elastic anisotropy are thus related by reversing the flows and rotating by $\pi$. In the case of $+1/2$ defects, for which the flow field is reversed by this rotation, this means that the active flow is identical for equal and opposite values of elastic anisotropy. This provides another demonstration that there cannot be a correction to the active flow of a $+1/2$ defect that is first order in $\kappa$. This symmetry of $+1/2$ defects is apparent in the top row of Figure \ref{fig:ActiveFlowsPerturbative}, in which the flows on the left match with the corresponding ones of the right. In particular, the pair with $\kappa=\pm 0.5$ corresponds to the case for which the orthogonality relation of the director was illustrated in Figure \ref{OrthogonalitySymmetry}a). 

\begin{figure}
    \centering
    \begin{tikzpicture}
        \node[anchor=south west,inner sep=0] at (0,0)
{\includegraphics[width=1\linewidth, trim = 0 0 0 0, clip, angle = 0, origin = c]{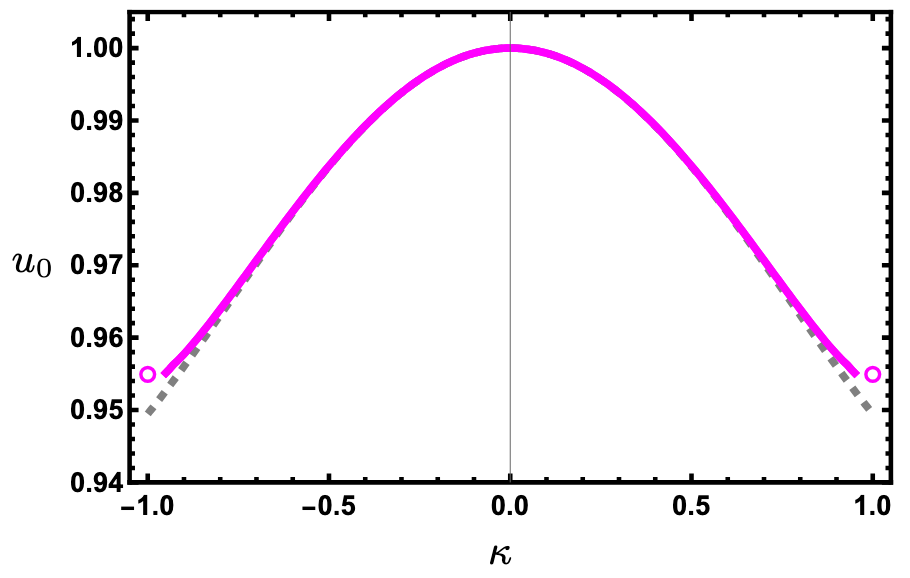}};

    \end{tikzpicture}
    \caption{The active motility, $u_0$, of a $+1/2$ defect as a function of elastic anisotropy, $\kappa$, normalised by the motility for equal elastic constants. The solid curve is the result of solving \eqref{eq:FirstIntegral} numerically and then integrating in \eqref{eq:Motility}, while the dashed curve comes from the quartic approximation shown in \eqref{eq:MotilityApproximation}. The open circles show the motility for maximal anisotropy, $\kappa=\pm1$, which can be found analytically to be $\frac{3}{\pi}$. The gap in the curve near $\kappa=\pm1$ is due to breakdown of convergence in the numerical solution of \eqref{eq:FirstIntegral}.}
    \label{ActiveMotilityPlusHalf}
\end{figure}

Restricting our focus to the flow at the location of a $+1/2$ defect gives the active motility shown in Figure~\ref{ActiveMotilityPlusHalf}. Setting $\mathbf{r}=0$ in \eqref{ActiveFlowIntegral} yields the following simple expression for the defect motility
\begin{equation}
    u_0=-\frac{\zeta R}{8\pi\mu}\int\cos(2\psi+\theta')+\frac{3}{2}\cos(2\psi-\theta')\text{d}\theta'.
    \label{eq:Motility}
\end{equation}
This shows that for isolated defects only those with index $s=+1/2$ or $s=+3/2$ can be actively motile \cite{tang2019theory}. This is natural in elastically isotropic systems, since these are the only topological defects with vectorial symmetry, and the same is true when there is elastic anisotropy, since this does not change the rotational symmetry of a defect. The motility of anisotropic defects can be found by solving \eqref{eq:FirstIntegral} numerically for $\psi$ and then numerically evaluating the integral in \eqref{eq:Motility}. When normalised by the isotropic motility, this produces the magenta curve shown in Figure \ref{ActiveMotilityPlusHalf}. That the active motility of the $+1/2$ defect is an even function of $\kappa$ is a direct consequence of the orthogonality relation on the director described earlier, with the vanishing of first-order contributions to the active flows manifest in the small change in magnitude, with Figure (\ref{ActiveMotilityPlusHalf}) showing that it never drops by more than 5\%. The grey dashed curve in Figure \ref{ActiveMotilityPlusHalf} is the result of solving \eqref{eq:FirstIntegral} for $\psi$ perturbatively to fourth order in $\kappa$, and then substituting into \eqref{eq:Motility}, giving the approximate motility
\begin{equation}
    u_0=-\frac{\zeta R}{4\mu}\left[1-\frac{9}{128}\kappa^2+\frac{81}{4096}\kappa^4+O(\kappa^6)\right].
    \label{eq:MotilityApproximation}
\end{equation}
Naturally, the first two terms accord with setting $r=0$ in \eqref{eq:ActiveFlowPlusHalf}.

\begin{figure}
    \centering
        \begin{tikzpicture}
            \node[anchor=south west,inner sep=0] at (0,0)
{\includegraphics[width=1\linewidth, trim = 0 0 0 0, clip]{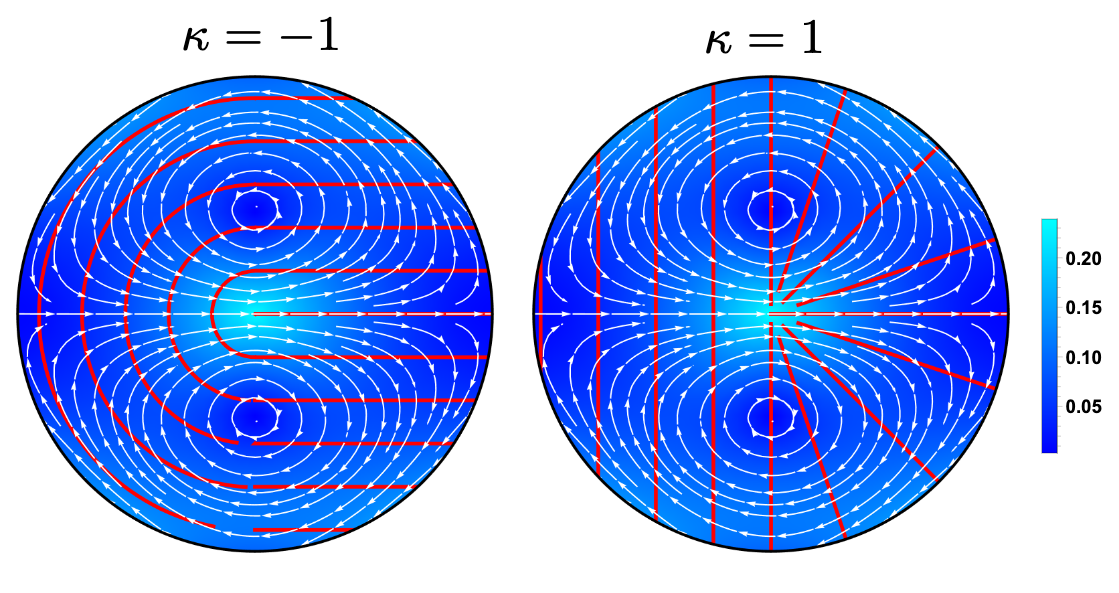}};
            
    \end{tikzpicture}
    \caption{The active flow around $+1/2$ defects with maximal elastic anisotropy. The streamlines of the flow are in white, superposed upon a background colour determined by the flow velocity, as indicated by the scale bar. The red lines indicate the director profile. These flows, given in \eqref{eq:ActiveFlowMaximalAnisotropy}, are derived from an exact solution for the director, rather than  the perturbative solution utilised in Figure \ref{fig:ActiveFlowsPerturbative}.
    }
    \label{fig:ActiveFlowPlusHalfDefectMaximalAnisotropy}
\end{figure}

The perturbative approach that we have taken relies on the series for the director angle, given in \eqref{eq:DirectorAnglePerturbative}, converging to the exact solution, but this is not true for $+1/2$ defects with maximal elastic anisotropy. In these cases the defect morphology consists exclusively of either splay or bend deformations and contains a line discontinuity in its gradient, not a feature that will be replicated by the perturbative expansion. 
Nonetheless the director field takes a simple form: a half-space of either pure bend or pure splay, complemented by a uniform director in the other half-space, as illustrated by the red lines in Figure~\ref{fig:ActiveFlowPlusHalfDefectMaximalAnisotropy}. This makes it possible to consider these profiles separately and use them as a point of comparison to test our earlier results.

The active flow field around a $+1/2$ defect with maximal elastic anisotropy is calculated in Appendix A.2, with the result given in \eqref{eq:ActiveFlowMaximalAnisotropy} and shown in Figure \ref{fig:ActiveFlowPlusHalfDefectMaximalAnisotropy}. Once again, opposite signs of anisotropy produce markedly distinct director profiles, but identical active flows, as a consequence of the symmetries described earlier. There is strong qualitative agreement between the flows depicted in Figure \ref{fig:ActiveFlowPlusHalfDefectMaximalAnisotropy} and the corresponding perturbative flow solutions in Figure \ref{fig:ActiveFlowsPerturbative}. Furthermore this exact calculation gives a normalised active motility of $\frac{3}{\pi}$, agreeing with the result of setting $\kappa=\pm1$ in \eqref{eq:MotilityApproximation} to within $0.6\%$, as can be seen in Figure \ref{ActiveMotilityPlusHalf}. Performing so well in the situation to which it is least-well adapted gives credence to our perturbative approach and suggests that it is not only capturing the qualitative behaviour of the system but also providing a good quantitative description.

\subsection{Defect pairs}
\label{subsec:DefectPairs}

Having established how elastic anisotropy modifies the active dynamics of isolated defects, we now turn our attention to its impact on their pairwise elastic interactions. We consider a $\pm 1/2$ defect pair in the infinite plane and find, via a calculation presented in Appendix B, that, to first order in the elastic anisotropy, the elastic energy is given by
\begin{equation}
    E=\pi\bar{K}\left[-\frac{1}{2}\ln\left(\frac{|\mathbf{r}_+-\mathbf{r}_-|}{a}\right)+\frac{\kappa}{3}\frac{\mathbf{p}\cdot(\mathbf{r}_+-\mathbf{r}_-)}{|\mathbf{r}_+-\mathbf{r}_-|}\right].
    \label{DefectPairEnergyElasticAnisotropy}
\end{equation}
Here the two defects are located at $\mathbf{r}_+$ and $\mathbf{r}_-$, $a$ is the defect core size and $\mathbf{p}$ is the orientation of the $+1/2$ defect, as defined in \eqref{DefectOrientation}. The first term of \eqref{DefectPairEnergyElasticAnisotropy} describes the standard Coulombic interaction between defects \cite{de1995physics}, with a strength set by the mean elastic constant. The second is a purely orientational contribution to the energy, giving rise to a torque, with a sign set by the nature of the elastic anisotropy. This contribution has been discussed previously \cite{ranganath1983energetics,chandrasekhar1986structure}, but without establishing its exact form. A torque of the same form has been found for $\pm 1/2$ defect pairs \cite{romano2024dynamical} via a different method than we present in Appendix B. We now elaborate upon the features of this anisotropic energy and its potential consequences in active systems.

\begin{figure}
    \centering
    \begin{tikzpicture}
        \node[anchor=south west,inner sep=0] at (0,0)
{\includegraphics[width=1\linewidth, trim = 0 0 0 0, clip]{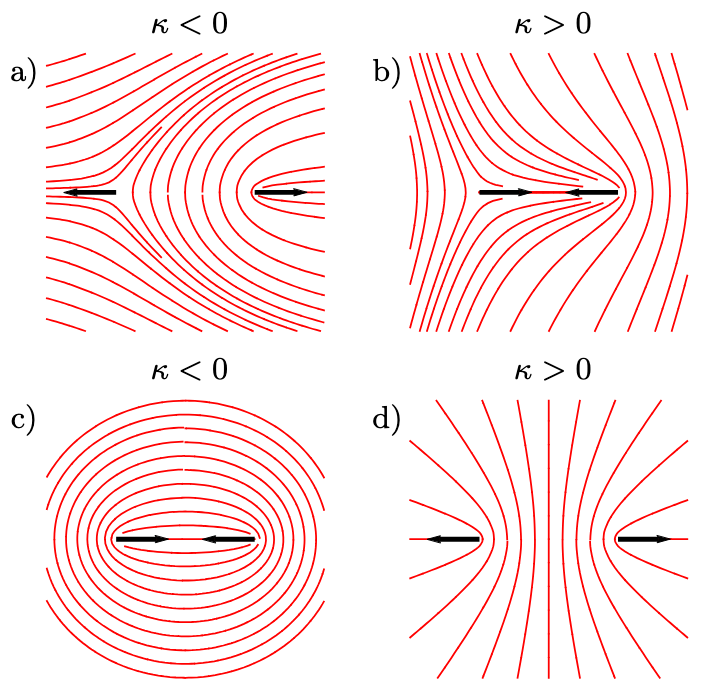}};
        
    \end{tikzpicture}
    \caption{The minimum energy configurations for defect pairs with elastic anisotropy. The red lines show the integral curves of the director and the arrows indicate the defect orientation, with a choice from the three possibilities having been made for the $-1/2$ defect. When $\kappa<0$, bend is favoured, causing oppositely charged defects to point away from each other a) and two $+1/2$ defects to face each other b). For $\kappa>0$, splay is promoted and the defect orientations are reversed.}
    \label{fig:DefectPairOrientation}
\end{figure}

We begin with the aspects that are common to all defect pairs, regardless of the defect charges. Firstly, varying the orientation of defects relative to their separation vector alters the amount of splay and bend in the system, and so elastic anisotropy induces a preference for a particular orientation, both defects pointing either towards or away from each other. For $\pm 1/2$ defects, $\kappa<0$ promotes the bend-dominated configuration shown in Figure \ref{fig:DefectPairOrientation}a) and $\kappa<0$ the largely splay distortion in Figure \ref{fig:DefectPairOrientation}b). Taking into account the changing direction of the far field, these defect configurations are near-field representations of the vertical and horizontal dipoles, and so their enhancement or suppression by elastic anisotropy mirrors that found for dipoles in \ref{subsubsec:2DActiveResponse}.

Secondly, changing the defect separation amounts to a rescaling of the system and so affects all director gradients equally. It is therefore not promoted by elastic anisotropy, the purpose of which is to bias the system towards splay or bend deformations. The energy due to elastic anisotropy is therefore purely orientational and independent of the distance between defects. Note that this scaling argument hinges upon the defect separation being the only relevant length and so cannot be applied for a finite system or when other defects are present, although an energy of the form of \eqref{DefectPairEnergyElasticAnisotropy} would still be expected as the dominant contribution in these cases. In particular, the distance-independence of the interaction does not mean that all defects interact with equal strength. Rather, the anisotropic elastic energy is dominated by a pairwise interaction between nearest neighbours, with a strength that is independent of the separation. This is because the mechanism of interaction is for defects to orient themselves so as to produce bend or splay deformations in the region between them, as illustrated in Figure \ref{fig:DefectPairOrientation}a) and b), a process that would be disrupted by interstitial defects.

Repeating the calculation for two $-1/2$ defects gives a similar torque as in equation \eqref{DefectPairEnergyElasticAnisotropy}, only half as strong. However, the analogous calculation breaks down for a pair of $+1/2$ defects, with the energetic preference for a given orientation becoming divergent. The reason for this can be seen from Figure \ref{fig:DefectPairOrientation}. The oppositely signed pairs shown in Figure \ref{fig:DefectPairOrientation}a) and b) have zero net charge, and so exist in an aligned far field, meaning the creation of bend or splay is largely confined to the region between the defects. On the other hand, the $+1/2$ defect pairs shown in Figure \ref{fig:DefectPairOrientation}c) and d) have the far-field character of a $+1$ defect, and hence can form infinite domains of pure bend or splay \cite{ranganath1983energetics,chandrasekhar1986structure}. This is a unique consequence of $+1$ defects being spin-$0$, that is lacking an intrinsic orientation \cite{tang2017orientation,houston2023active}. Since the far field lacks an orientation, it is instead its character that alters, varying from the bend-dominated vortex in Figure \ref{fig:DefectPairOrientation}c) to the splay-dominated aster in Figure \ref{fig:DefectPairOrientation}d). Regularising the divergence through a large-scale cut-off $R$ gives an anisotropic elastic energy $\sim\bar{K}\kappa\log R$. In a multi-defect system this cut-off could be provided by a third defect, the presence of which halts the bubble of pure bend or splay. This suggests that pairs of $+1/2$ defects have a nearest neighbour interaction that scales logarithmically with the distance to the next-nearest neighbour. In the context of active nematics, it is interesting to consider how this interaction can therefore depend on activity through both the effective elastic anisotropy \cite{kumar2018tunable} and the typical defect separation \cite{thampi2014vorticity,giomi2014defect}.

More generally, the torque we have identified, although passive in origin, can nonetheless affect the dynamics of active nematic defects by influencing whether activity works with or against the elastic forces. It is clear from Figure \ref{fig:DefectPairOrientation}a) that $\kappa<0$ promotes defect annihilation in extensile systems and favours separation in contractile ones, with the converse holding when $\kappa>0$, as in Figure \ref{fig:DefectPairOrientation}b). In situations where the elastic and active forces oppose one another a balance between the two may arise, leading to metastable states where the defect separation remains constant over long timescales \cite{kumar2018tunable}. Even randomising torques are able to confound the directed motion of $+1/2$ defects away from their $-1/2$ counterparts thus enhancing the nematic ordering, as has been demonstrated in the context of thermal fluctuations \cite{shankar2018defect}. The torques produced through elastic anisotropy would modify the activity-temperature phase diagram governing the transition between the nematic and isotropic states. In particular, with the appropriate sign of elastic anisotropy, there would be a temperature-independent torque which would act so as to remove the re-entrant isotropic phase transition in the limit of low activity and temperature described in \cite{shankar2018defect}. Lastly, the elastic torque in \eqref{DefectPairEnergyElasticAnisotropy} could modify the trajectories of defects during creation and annihilation events, such that they travel along spiral paths, as has been found to result from incorporating polar active forces in simulations of active nematics \cite{beer2025irreversibility}.

It is worth distinguishing the elastic torque we have described from that which can arise in elastically isotropic systems \cite{vromans2016orientational,tang2017orientation,vcopar2024many}. Such torques stem from the orientation of two defects being allowed to vary independently, such that they may become misaligned. In \cite{tang2017orientation} this is achieved by a conformal mapping, resulting in a torque that decays logarithmically with defect separation. In \cite{vromans2016orientational} the appropriate director field is attained via an image system, while in \cite{vcopar2024many} each defect is imbued with a geometric `spiral charge', with both methods producing a torque independent of defect separation. In our work the orientations of the two defects are locked together and varied relative to their separation vector. All such configurations have the same energy with equal elastic constants, the degeneracy only broken by elastic anisotropy. Since it is independent of defect separation, this anisotropic elastic torque is comparable in magnitude to the isotropic one, and indeed for the case of two $+1/2$ defects may dominate it.

\section{Discussion}
We have considered the effect of elastic anisotropy on the dynamics of active nematics, via the active response induced by generic distortions, along with the active flows and elastic energies of topological defects in two dimensions. Under a one-elastic-constant approximation, certain dipole distortions induce active forces, while torques are associated with quadrupoles \cite{houston2023active,houston2022defect,houston2023colloids}. The chief departure from this picture is that elastic anisotropy leads to novel active torques associated with monopole distortions. We show that these new torques arise because elastic anisotropy induces a perturbation to the harmonic monopole which has quadrupolar symmetry. This modification of the fundamental active response leads to the active forces and torques induced by different dipoles and quadrupoles no longer being equal, breaking, for example, the standard relation between distortions and active propulsion direction. Indeed, we find that in three dimensions the effect of elastic anisotropy can be sufficient to reverse the defect-driven propulsion of a colloid.

Turning, in two dimensions, to topological defects themselves, we find that the symmetry of $+1/2$ defects renders their active flows uniquely robust to variations in elasticity, despite the large variation it causes in their morphology. However, for defect pairs elastic anisotropy introduces elastic torques, which we expect to be comparable in magnitude to those that arise under a one-elastic-constant approximation.

In this work we identified new phenomena that arise due to elastic anisotropy, but have not explored the parallel story that exists for effects resulting from anisotropy of the viscous response. This viscous anisotropy could be incorporated in a similar fashion to our treatment of elastic anisotropy, that is by expanding the flow response to first order in deviations from isotropic viscosity and decomposing into irreducible representations under rotations about the far-field director. This would imbue the fundamental monopole distortion with additional active responses of various spin, which would again filter down to higher-order multipole distortions via differentiation. Such a combined treatment of anisotropic elasticity and viscosity via irreducible representations has revealed odd mechanics in active columnar phases \cite{kole2024chirality}. The perturbative analysis of distinct viscosities has been performed for singular Stokes flows in a nematic environment \cite{kos2018elementary}, showing that there is a misalignment between the point force and the flow response whenever the force is not along or perpendicular to the nematic director. This mirrors our results on the propulsive force induced by a dipolar distortion with elastic anisotropy, and in this context we have shown that similar effects arise due to viscous anisotropy, with the two forms of anisotropy able to combine constructively or destructively.

Elastic anisotropy leads to a number of important effects in passive nematics \cite{lavrentovich2024splay} and our results indicate the same will be true in their active analogues. There have already been indications of this in the creation of novel dynamic states through banded elasticity \cite{kumar2022catapulting} and the modification of stress patterns around topological defects \cite{bonn2024elasticity}. By showing that elastic anisotropy leads to the association of active torques with monopole distortions our work opens up a new pathway for rotational active dynamics, including ratchet effects \cite{di2010bacterial,reichhardt2017ratchet,ray2023rectified,houston2023colloids}. There is the potential for this effect to be dominant in certain circumstances, since monopoles decay much slower than quadrupoles. It would be interesting to incorporate the anisotropy-induced elastic torques we have described into descriptions of multi-defect dynamics \cite{vafa2020multi}, to determine if they play a role in the potential long-range ordering of defects \cite{decamp2015orientational,thijssen2020role,pearce2021orientational}. Similar elastic torques could play a role in reorientating defect lines, such that splay-bend anisotropy accelerates the twist-induced crossover from quasi-two-dimensional to fully three-dimensional defect dynamics \cite{shendruk2018twist}.

\section*{Acknowledgements}
I am very grateful to Gareth Alexander for support and guidance throughout this work, as well as comments on the manuscript. This work was supported by the UK EPSRC through Grant No. EP/N509796/1.

\bibliography{ElasticAnisotropy.bib}

@article{di2008hydrodynamic,
  title={Hydrodynamic interactions in two dimensions},
  author={Di Leonardo, R and Keen, S and Ianni, F and Leach, J and Padgett, M.J. and Ruocco, G},
  journal={Physical Review E},
  volume={78},
  number={3},
  pages={031406},
  year={2008},
  publisher={APS}
}

@article{giomi2014defect,
  title={Defect dynamics in active nematics},
  author={Giomi, Luca and Bowick, M.J. and Mishra, Prashant and Sknepnek, Rastko and Marchetti, M.C.},
  journal={Philosophical Transactions of the Royal Society A: Mathematical, Physical and Engineering Sciences},
  volume={372},
  number={2029},
  pages={20130365},
  year={2014},
  publisher={The Royal Society Publishing}
}

@article{tang2019theory,
  title={Theory of defect motion in 2D passive and active nematic liquid crystals},
  author={Tang, Xingzhou and Selinger, J.V.},
  journal={Soft matter},
  volume={15},
  number={4},
  pages={587--601},
  year={2019},
  publisher={Royal Society of Chemistry}
}

@article{tang2017orientation,
  title={Orientation of topological defects in 2D nematic liquid crystals},
  author={Tang, Xingzhou and Selinger, J.V.},
  journal={Soft Matter},
  volume={13},
  number={32},
  pages={5481--5490},
  year={2017},
  publisher={Royal Society of Chemistry}
}

@article{kumar2018tunable,
  title={Tunable structure and dynamics of active liquid crystals},
  author={Kumar, Nitin and Zhang, Rui and de Pablo, J.J. and Gardel, M.L.},
  journal={Science advances},
  volume={4},
  number={10},
  pages={eaat7779},
  year={2018},
  publisher={American Association for the Advancement of Science}
}

@article{zhang2018interplay,
  title={Interplay of structure, elasticity, and dynamics in actin-based nematic materials},
  author={Zhang, Rui and Kumar, Nitin and Ross, J.L. and Gardel, M.L. and de Pablo, J.J.},
  journal={Proceedings of the National Academy of Sciences},
  volume={115},
  number={2},
  pages={E124--E133},
  year={2018},
  publisher={National Acad Sciences}
}

@book{landau1995course,
  title={Course of Theoretical Physics: Theory of Elasticity},
  author={Landau, L.D. and Lifshitz, E.M. and Berestetskii, V.B. and Pitaevskii, L.P.},
  year={1995}
}

@article{simha2002hydrodynamic,
  title={Hydrodynamic fluctuations and instabilities in ordered suspensions of self-propelled particles},
  author={Simha, R.A. and Ramaswamy, Sriram},
  journal={Physical review letters},
  volume={89},
  number={5},
  pages={058101},
  year={2002},
  publisher={APS}
}

@article{ramaswamy2010mechanics,
  title={The mechanics and statistics of active matter},
  author={Ramaswamy, Sriram},
  journal={Annu. Rev. Condens. Matter Phys.},
  volume={1},
  number={1},
  pages={323--345},
  year={2010},
  publisher={Annual Reviews}
}

@article{marchetti2013hydrodynamics,
  title={Hydrodynamics of soft active matter},
  author={Marchetti, M.C. and Joanny, Jean-Fran{\c{c}}ois and Ramaswamy, Sriram and Liverpool, Tanniemola B and Prost, Jacques and Rao, Madan and Simha, RA},
  journal={Reviews of Modern Physics},
  volume={85},
  number={3},
  pages={1143},
  year={2013},
  publisher={APS}
}

@article{doostmohammadi2018active,
  title={Active nematics},
  author={Doostmohammadi, Amin and Ign{\'e}s-Mullol, Jordi and Yeomans, J.M. and Sagu{\'e}s, Francesc},
  journal={Nature communications},
  volume={9},
  number={1},
  pages={3246},
  year={2018},
  publisher={Nature Publishing Group}
}

@article{sanchez2012spontaneous,
  title={Spontaneous motion in hierarchically assembled active matter},
  author={Sanchez, Tim and Chen, D.T.N. and DeCamp, S.J. and Heymann, Michael and Dogic, Zvonimir},
  journal={Nature},
  volume={491},
  number={7424},
  pages={431},
  year={2012},
  publisher={Nature Publishing Group}
}

@article{vromans2016orientational,
  title={Orientational properties of nematic disclinations},
  author={Vromans, A.J. and Giomi, Luca},
  journal={Soft matter},
  volume={12},
  number={30},
  pages={6490--6495},
  year={2016},
  publisher={Royal Society of Chemistry}
}

@article{shankar2018defect,
  title={Defect unbinding in active nematics},
  author={Shankar, Suraj and Ramaswamy, Sriram and Marchetti, M.C. and Bowick, M.J.},
  journal={Physical review letters},
  volume={121},
  number={10},
  pages={108002},
  year={2018},
  publisher={APS}
}

@article{giomi2013defect,
  title={Defect annihilation and proliferation in active nematics},
  author={Giomi, Luca and Bowick, Mark J and Ma, Xu and Marchetti, M Cristina},
  journal={Physical review letters},
  volume={110},
  number={22},
  pages={228101},
  year={2013},
  publisher={APS}
}

@article{thampi2014vorticity,
  title={Vorticity, defects and correlations in active turbulence},
  author={Thampi, Sumesh P and Golestanian, Ramin and Yeomans, Julia M},
  journal={Philosophical Transactions of the Royal Society A: Mathematical, Physical and Engineering Sciences},
  volume={372},
  number={2029},
  pages={20130366},
  year={2014},
  publisher={The Royal Society Publishing}
}

@book{de1995physics,
  title={The physics of liquid crystals},
  author={De Gennes, Pierre-Gilles and Prost, Jacques},
  volume={83},
  year={1995},
  publisher={Oxford university press}
}

@article{voituriez2005spontaneous,
  title={Spontaneous flow transition in active polar gels},
  author={Voituriez, R and Joanny, Jean-Fran{\c{c}}ois and Prost, Jacques},
  journal={EPL (Europhysics Letters)},
  volume={70},
  number={3},
  pages={404},
  year={2005},
  publisher={IOP Publishing}
}

@article{edwards2009spontaneous,
  title={Spontaneous flow states in active nematics: a unified picture},
  author={Edwards, SA and Yeomans, JM},
  journal={EPL (Europhysics Letters)},
  volume={85},
  number={1},
  pages={18008},
  year={2009},
  publisher={IOP Publishing}
}

@article{dzyaloshinskii1970theory,
  title={Theory of disinclinations in liquid crystals},
  author={Dzyaloshinskii, IE},
  journal={Soviet Journal of Experimental and Theoretical Physics},
  volume={31},
  pages={773},
  year={1970}
}

@article{ranganath1983energetics,
  title={Energetics of disclinations in liquid crystals},
  author={Ranganath, GS},
  journal={Molecular Crystals and Liquid Crystals},
  volume={97},
  number={1},
  pages={77--94},
  year={1983},
  publisher={Taylor \& Francis}
}

@article{hudson1989frank,
  title={Frank elastic-constant anisotropy measured from transmission-electron-microscope images of disclinations},
  author={Hudson, Steven D and Thomas, Edwin L},
  journal={Physical review letters},
  volume={62},
  number={17},
  pages={1993},
  year={1989},
  publisher={APS}
}

@article{chandrasekhar1986structure,
  title={The structure and energetics of defects in liquid crystals},
  author={Chandrasekhar, S and Ranganath, GS},
  journal={Advances in Physics},
  volume={35},
  number={6},
  pages={507--596},
  year={1986},
  publisher={Taylor \& Francis}
}

@article{joshi2019interplay,
  title={The interplay between activity and filament flexibility determines the emergent properties of active nematics},
  author={Joshi, Abhijeet and Putzig, Elias and Baskaran, Aparna and Hagan, Michael F},
  journal={Soft matter},
  volume={15},
  number={1},
  pages={94--101},
  year={2019},
  publisher={Royal Society of Chemistry}
}

@article{shendruk2018twist,
  title={Twist-induced crossover from two-dimensional to three-dimensional turbulence in active nematics},
  author={Shendruk, Tyler N and Thijssen, Kristian and Yeomans, Julia M and Doostmohammadi, Amin},
  journal={Physical Review E},
  volume={98},
  number={1},
  pages={010601},
  year={2018},
  publisher={APS}
}

@article{saw2017topological,
  title={Topological defects in epithelia govern cell death and extrusion},
  author={Saw, Thuan Beng and Doostmohammadi, Amin and Nier, Vincent and Kocgozlu, Leyla and Thampi, Sumesh and Toyama, Yusuke and Marcq, Philippe and Lim, Chwee Teck and Yeomans, Julia M and Ladoux, Benoit},
  journal={Nature},
  volume={544},
  number={7649},
  pages={212},
  year={2017},
  publisher={Nature Publishing Group}
}

@article{binysh2019three,
  title={Three-Dimensional Active Defect Loops},
  author={Binysh, Jack and Kos, {\v{Z}}iga and {\v{C}}opar, Simon and Ravnik, Miha and Alexander, Gareth P},
  journal={arXiv preprint arXiv:1909.07109},
  year={2019}
}

@article{duclos2017topological,
  title={Topological defects in confined populations of spindle-shaped cells},
  author={Duclos, Guillaume and Erlenk{\"a}mper, Christoph and Joanny, Jean-Fran{\c{c}}ois and Silberzan, Pascal},
  journal={Nature Physics},
  volume={13},
  number={1},
  pages={58},
  year={2017},
  publisher={Nature Publishing Group}
}

@article{shendruk2017dancing,
  title={Dancing disclinations in confined active nematics},
  author={Shendruk, Tyler N and Doostmohammadi, Amin and Thijssen, Kristian and Yeomans, Julia M},
  journal={Soft Matter},
  volume={13},
  number={21},
  pages={3853--3862},
  year={2017},
  publisher={Royal Society of Chemistry}
}

@article{giomi2015geometry,
  title={Geometry and topology of turbulence in active nematics},
  author={Giomi, Luca},
  journal={Physical Review X},
  volume={5},
  number={3},
  pages={031003},
  year={2015},
  publisher={APS}
}

@article{houston2022defect,
  title={Defect loops in three-dimensional active nematics as active multipoles},
  author={Houston, Alexander J H and Alexander, Gareth P},
  journal={Physical Review E},
  volume={105},
  number={6},
  pages={L062601},
  year={2022},
  publisher={APS}
}

@article{houston2023colloids,
  title={Colloids in two-dimensional active nematics: conformal cogs and controllable spontaneous rotation},
  author={Houston, Alexander J H and Alexander, Gareth P},
  journal={New Journal of Physics},
  volume={25},
  number={12},
  pages={123006},
  year={2023},
  publisher={IOP Publishing}
}

@article{houston2023active,
  title={Active nematic multipoles: flow responses and the dynamics of defects and colloids},
  author={Houston, Alexander J H and Alexander, Gareth P},
  journal={Frontiers in Physics},
  volume={11},
  pages={1110244},
  year={2023},
  publisher={Frontiers Media SA}
}

@article{ray2023rectified,
  title={Rectified rotational dynamics of mobile inclusions in two-dimensional active nematics},
  author={Ray, Sattvic and Zhang, Jie and Dogic, Zvonimir},
  journal={Physical Review Letters},
  volume={130},
  number={23},
  pages={238301},
  year={2023},
  publisher={APS}
}

@article{angheluta2021role,
  title={The role of fluid flow in the dynamics of active nematic defects},
  author={Angheluta, Luiza and Chen, Zhitao and Marchetti, M Cristina and Bowick, Mark J},
  journal={New Journal of Physics},
  volume={23},
  number={3},
  pages={033009},
  year={2021},
  publisher={IOP Publishing}
}

@article{khoromskaia2017vortex,
  title={Vortex formation and dynamics of defects in active nematic shells},
  author={Khoromskaia, Diana and Alexander, Gareth P},
  journal={New Journal of Physics},
  volume={19},
  number={10},
  pages={103043},
  year={2017},
  publisher={IOP Publishing}
}

@article{alert2020universal,
  title={Universal scaling of active nematic turbulence},
  author={Alert, Ricard and Joanny, Jean-Fran{\c{c}}ois and Casademunt, Jaume},
  journal={Nature Physics},
  volume={16},
  number={6},
  pages={682--688},
  year={2020},
  publisher={Nature Publishing Group UK London}
}

@article{romano2024dynamical,
  title={Dynamical theory of topological defects II: universal aspects of defect motion},
  author={Romano, Jacopo and Mahault, Beno{\^\i}t and Golestanian, Ramin},
  journal={Journal of Statistical Mechanics: Theory and Experiment},
  volume={2024},
  number={3},
  pages={033208},
  year={2024},
  publisher={IOP Publishing}
}

@article{vcopar2024many,
  title={Many-defect solutions in planar nematics: interactions, spiral textures and boundary conditions},
  author={{\v{C}}opar, Simon and Kos, {\v{Z}}iga},
  journal={Soft Matter},
  volume={20},
  number={35},
  pages={6859–7084},
  year={2024}
}

@article{boocock2023interplay,
  title={Interplay between mechanochemical patterning and glassy dynamics in cellular monolayers},
  author={Boocock, Daniel and Hirashima, Tsuyoshi and Hannezo, Edouard},
  journal={PRX Life},
  volume={1},
  number={1},
  pages={013001},
  year={2023},
  publisher={APS}
}

@article{brugues2014forces,
  title={Forces driving epithelial wound healing},
  author={Brugu{\'e}s, Agust{\'\i} and Anon, Ester and Conte, Vito and Veldhuis, Jim H and Gupta, Mukund and Colombelli, Julien and Mu{\~n}oz, Jos{\'e} J and Brodland, G Wayne and Ladoux, Benoit and Trepat, Xavier},
  journal={Nature physics},
  volume={10},
  number={9},
  pages={683--690},
  year={2014},
  publisher={Nature Publishing Group UK London}
}

@book{ahlfors2006lectures,
  title={Lectures on quasiconformal mappings},
  author={Ahlfors, Lars Valerian},
  volume={38},
  year={2006},
  publisher={American Mathematical Soc.}
}

@article{wensink2012meso,
  title={Meso-scale turbulence in living fluids},
  author={Wensink, Henricus H and Dunkel, J{\"o}rn and Heidenreich, Sebastian and Drescher, Knut and Goldstein, Raymond E and L{\"o}wen, Hartmut and Yeomans, Julia M},
  journal={Proceedings of the Proceedings of the National Academy of Sciences},
  volume={109},
  number={36},
  pages={14308--14313},
  year={2012},
  publisher={National Academy Sciences}
}

@article{zhou2014living,
  title={Living liquid crystals},
  author={Zhou, Shuang and Sokolov, Andrey and Lavrentovich, Oleg D and Aranson, Igor S},
  journal={Biophysical Journal},
  volume={106},
  number={2},
  pages={420a},
  year={2014},
  publisher={Elsevier}
}

@article{aditi2002hydrodynamic,
  title={Hydrodynamic fluctuations and instabilities in ordered suspensions of self-propelled particles},
  author={Aditi Simha, R and Ramaswamy, Sriram},
  journal={Physical Review Letters},
  volume={89},
  number={5},
  pages={058101},
  year={2002},
  publisher={APS}
}

@article{maroudas2021topological,
  title={Topological defects in the nematic order of actin fibres as organization centres of Hydra morphogenesis},
  author={Maroudas-Sacks, Yonit and Garion, Liora and Shani-Zerbib, Lital and Livshits, Anton and Braun, Erez and Keren, Kinneret},
  journal={Nature Physics},
  volume={17},
  number={2},
  pages={251--259},
  year={2021},
  publisher={Nature Publishing Group UK London}
}

@article{zhang2021spatiotemporal,
  title={Spatiotemporal control of liquid crystal structure and dynamics through activity patterning},
  author={Zhang, Rui and Redford, Steven A and Ruijgrok, Paul V and Kumar, Nitin and Mozaffari, Ali and Zemsky, Sasha and Dinner, Aaron R and Vitelli, Vincenzo and Bryant, Zev and Gardel, Margaret L and  de Pablo, Juan J. },
  journal={Nature Materials},
  volume={20},
  number={6},
  pages={875--882},
  year={2021},
  publisher={Nature Publishing Group UK London}
}

@article{shankar2024design,
  title={Design rules for controlling active topological defects},
  author={Shankar, Suraj and Scharrer, Luca VD and Bowick, Mark J and Marchetti, M Cristina},
  journal={Proceedings of the National Academy of Sciences},
  volume={121},
  number={21},
  pages={e2400933121},
  year={2024},
  publisher={National Academy Sciences}
}

@article{partovifard2024controlling,
  title={Controlling active turbulence by activity patterns},
  author={Partovifard, Arghavan and Grawitter, Josua and Stark, Holger},
  journal={Soft Matter},
  volume={20},
  number={8},
  pages={1800--1814},
  year={2024},
  publisher={Royal Society of Chemistry}
}

@article{schimming2025turbulence,
  title={Turbulence-to-order transitions in activity-patterned active nematics},
  author={Schimming, Cody D and Reichhardt, CJO and Reichhardt, C},
  journal={Physical Review E},
  volume={111},
  number={3},
  pages={035404},
  year={2025},
  publisher={APS}
}

@article{houston2024spontaneous,
  title={Spontaneous flows and quantum analogies in heterogeneous active nematic films},
  author={Houston, Alexander J H and Mottram, Nigel J},
  journal={Communications Physics},
  volume={7},
  number={1},
  pages={375},
  year={2024},
  publisher={Nature Publishing Group UK London}
}

@article{vining2017mechanical,
  title={Mechanical forces direct stem cell behaviour in development and regeneration},
  author={Vining, Kyle H and Mooney, David J},
  journal={Nature reviews Molecular cell biology},
  volume={18},
  number={12},
  pages={728--742},
  year={2017},
  publisher={Nature Publishing Group UK London}
}

@article{ladoux2017mechanobiology,
  title={Mechanobiology of collective cell behaviours},
  author={Ladoux, Benoit and M{\`e}ge, Ren{\'e}-Marc},
  journal={Nature reviews Molecular cell biology},
  volume={18},
  number={12},
  pages={743--757},
  year={2017},
  publisher={Nature Publishing Group UK London}
}

@article{hayward2021tissue,
  title={Tissue mechanics in stem cell fate, development, and cancer},
  author={Hayward, Mary-Kate and Muncie, Jonathon M and Weaver, Valerie M},
  journal={Developmental cell},
  volume={56},
  number={13},
  pages={1833--1847},
  year={2021},
  publisher={Elsevier}
}

@article{persat2015mechanical,
  title={The mechanical world of bacteria},
  author={Persat, Alexandre and Nadell, Carey D and Kim, Minyoung Kevin and Ingremeau, Francois and Siryaporn, Albert and Drescher, Knut and Wingreen, Ned S and Bassler, Bonnie L and Gitai, Zemer and Stone, Howard A},
  journal={Cell},
  volume={161},
  number={5},
  pages={988--997},
  year={2015},
  publisher={Elsevier}
}

@article{marenduzzo2007steady,
  title={Steady-state hydrodynamic instabilities of active liquid crystals: Hybrid lattice Boltzmann simulations},
  author={Marenduzzo, D and Orlandini, Enzo and Cates, ME and Yeomans, JM},
  journal={Physical Review E—Statistical, Nonlinear, and Soft Matter Physics},
  volume={76},
  number={3},
  pages={031921},
  year={2007},
  publisher={APS}
}

@article{duclos2018spontaneous,
  title={Spontaneous shear flow in confined cellular nematics},
  author={Duclos, G and Blanch-Mercader, C and Yashunsky, V and Salbreux, G and Joanny, J-F and Prost, J and Silberzan, Pascal},
  journal={Nature physics},
  volume={14},
  number={7},
  pages={728--732},
  year={2018},
  publisher={Nature Publishing Group UK London}
}

@article{thampi2016active,
  title={Active turbulence in active nematics},
  author={Thampi, SumeshP and Yeomans, JuliaM},
  journal={The European Physical Journal Special Topics},
  volume={225},
  pages={651--662},
  year={2016},
  publisher={Springer}
}

@article{doostmohammadi2017onset,
  title={Onset of meso-scale turbulence in active nematics},
  author={Doostmohammadi, Amin and Shendruk, Tyler N and Thijssen, Kristian and Yeomans, Julia M},
  journal={Nature communications},
  volume={8},
  number={1},
  pages={15326},
  year={2017},
  publisher={Nature Publishing Group UK London}
}

@article{alert2022active,
  title={Active turbulence},
  author={Alert, Ricard and Casademunt, Jaume and Joanny, Jean-Fran{\c{c}}ois},
  journal={Annual Review of Condensed Matter Physics},
  volume={13},
  number={1},
  pages={143--170},
  year={2022},
  publisher={Annual Reviews}
}

@article{gudipaty2017mechanical,
  title={Mechanical stretch triggers rapid epithelial cell division through Piezo1},
  author={Gudipaty, Swapna A and Lindblom, Jody and Loftus, Patrick D and Redd, Michael J and Edes, Kornelia and Davey, CF and Krishnegowda, V and Rosenblatt, Jody},
  journal={Nature},
  volume={543},
  number={7643},
  pages={118--121},
  year={2017},
  publisher={Nature Publishing Group UK London}
}

@article{eisenhoffer2012crowding,
  title={Crowding induces live cell extrusion to maintain homeostatic cell numbers in epithelia},
  author={Eisenhoffer, George T and Loftus, Patrick D and Yoshigi, Masaaki and Otsuna, Hideo and Chien, Chi-Bin and Morcos, Paul A and Rosenblatt, Jody},
  journal={Nature},
  volume={484},
  number={7395},
  pages={546--549},
  year={2012},
  publisher={Nature Publishing Group UK London}
}

@article{le2024mechanical,
  title={Mechanical stresses govern myoblast fusion and myotube growth},
  author={Le Toquin, Yoann and Dubey, Sushil and Arda{\v{s}}eva, Aleksandra and Balasubramaniam, Lakshmi and Delaune, Emilie and Morin, Val{\'e}rie and Doostmohammadi, Amin and Marcelle, Christophe and Ladoux, Beno{\^\i}t},
  journal={bioRxiv},
  pages={2024--11},
  year={2024},
  publisher={Cold Spring Harbor Laboratory}
}

@article{saw2018biological,
  title={Biological tissues as active nematic liquid crystals},
  author={Saw, Thuan Beng and Xi, Wang and Ladoux, Benoit and Lim, Chwee Teck},
  journal={Advanced materials},
  volume={30},
  number={47},
  pages={1802579},
  year={2018},
  publisher={Wiley Online Library}
}

@article{zhang2021autonomous,
  title={Autonomous materials systems from active liquid crystals},
  author={Zhang, Rui and Mozaffari, Ali and de Pablo, Juan J},
  journal={Nature Reviews Materials},
  volume={6},
  number={5},
  pages={437--453},
  year={2021},
  publisher={Nature Publishing Group UK London}
}

@article{copenhagen2021topological,
  title={Topological defects promote layer formation in Myxococcus xanthus colonies},
  author={Copenhagen, Katherine and Alert, Ricard and Wingreen, Ned S and Shaevitz, Joshua W},
  journal={Nature Physics},
  volume={17},
  number={2},
  pages={211--215},
  year={2021},
  publisher={Nature Publishing Group UK London}
}

@article{kumar2022catapulting,
  title={Catapulting of topological defects through elasticity bands in active nematics},
  author={Kumar, Nitin and Zhang, Rui and Redford, Steven A and de Pablo, Juan J and Gardel, Margaret L},
  journal={Soft Matter},
  volume={18},
  number={28},
  pages={5271--5281},
  year={2022},
  publisher={Royal Society of Chemistry}
}

@article{lavrentovich2024splay,
  title={Splay-bend elastic inequalities shape tactoids, toroids, umbilics, and conic section walls in paraelectric, twist-bend, and ferroelectric nematics},
  author={Lavrentovich, Oleg D},
  journal={Liquid Crystals Reviews},
  volume={12},
  number={1},
  pages={1--13},
  year={2024},
  publisher={Taylor \& Francis}
}

@article{bonn2024elasticity,
  title={Elasticity tunes mechanical stress localization around active topological defects},
  author={Bonn, Lasse and Arda{\v{s}}eva, Aleksandra and Doostmohammadi, Amin},
  journal={Soft Matter},
  volume={20},
  number={1},
  pages={115--123},
  year={2024},
  publisher={Royal Society of Chemistry}
}

@article{norton2018insensitivity,
  title={Insensitivity of active nematic liquid crystal dynamics to topological constraints},
  author={Norton, Michael M and Baskaran, Arvind and Opathalage, Achini and Langeslay, Blake and Fraden, Seth and Baskaran, Aparna and Hagan, Michael F},
  journal={Physical Review E},
  volume={97},
  number={1},
  pages={012702},
  year={2018},
  publisher={APS}
}

@article{opathalage2019self,
  title={Self-organized dynamics and the transition to turbulence of confined active nematics},
  author={Opathalage, Achini and Norton, Michael M and Juniper, Michael PN and Langeslay, Blake and Aghvami, S Ali and Fraden, Seth and Dogic, Zvonimir},
  journal={Proceedings of the National Academy of Sciences},
  volume={116},
  number={11},
  pages={4788--4797},
  year={2019},
  publisher={National Academy of Sciences}
}

@article{di2010bacterial,
  title={Bacterial ratchet motors},
  author={Di Leonardo, Roberto and Angelani, Luca and Dell’Arciprete, Dario and Ruocco, Giancarlo and Iebba, Valerio and Schippa, Serena and Conte, Maria Pia and Mecarini, Francesco and De Angelis, Francesco and Di Fabrizio, Enzo},
  journal={Proceedings of the National Academy of Sciences},
  volume={107},
  number={21},
  pages={9541--9545},
  year={2010},
  publisher={National Academy of Sciences}
}

@article{vafa2020multi,
  title={Multi-defect dynamics in active nematics},
  author={Vafa, Farzan and Bowick, Mark J and Marchetti, M Cristina and Shraiman, Boris I},
  journal={arXiv preprint arXiv:2007.02947},
  year={2020}
}

@article{decamp2015orientational,
  title={Orientational order of motile defects in active nematics},
  author={DeCamp, Stephen J and Redner, Gabriel S and Baskaran, Aparna and Hagan, Michael F and Dogic, Zvonimir},
  journal={Nature materials},
  volume={14},
  number={11},
  pages={1110--1115},
  year={2015},
  publisher={Nature Publishing Group UK London}
}

@article{reichhardt2017ratchet,
  title={Ratchet effects in active matter systems},
  author={Reichhardt, CJ Olson and Reichhardt, Charles},
  journal={Annual Review of Condensed Matter Physics},
  volume={8},
  number={1},
  pages={51--75},
  year={2017},
  publisher={Annual Reviews}
}

@article{thijssen2020role,
  title={Role of friction in multidefect ordering},
  author={Thijssen, Kristian and Nejad, Mehrana R and Yeomans, Julia M},
  journal={Physical Review Letters},
  volume={125},
  number={21},
  pages={218004},
  year={2020},
  publisher={APS}
}

@article{pearce2021orientational,
  title={Orientational correlations in active and passive nematic defects},
  author={Pearce, DJG and Nambisan, J and Ellis, PW and Fernandez-Nieves, A and Giomi, L},
  journal={Physical Review Letters},
  volume={127},
  number={19},
  pages={197801},
  year={2021},
  publisher={APS}
}

@article{chandrakar2020confinement,
  title={Confinement controls the bend instability of three-dimensional active liquid crystals},
  author={Chandrakar, Pooja and Varghese, Minu and Aghvami, S Ali and Baskaran, Aparna and Dogic, Zvonimir and Duclos, Guillaume},
  journal={Physical review letters},
  volume={125},
  number={25},
  pages={257801},
  year={2020},
  publisher={APS}
}

@article{kole2024chirality,
  title={Chirality and odd mechanics in active columnar phases},
  author={Kole, SJ and Alexander, Gareth P and Maitra, Ananyo and Ramaswamy, Sriram},
  journal={PNAS nexus},
  volume={3},
  number={10},
  pages={pgae398},
  year={2024},
  publisher={Oxford University Press US}
}

@article{kos2018elementary,
  title={Elementary flow field profiles of micro-swimmers in weakly anisotropic nematic fluids: Stokeslet, stresslet, rotlet and source flows},
  author={Kos, {\v{Z}}iga and Ravnik, Miha},
  journal={Fluids},
  volume={3},
  number={1},
  pages={15},
  year={2018},
  publisher={MDPI}
}

@book{kim2013microhydrodynamics,
  title={Microhydrodynamics: principles and selected applications},
  author={Kim, Sangtae and Karrila, Seppo J},
  year={2013},
  publisher={Butterworth-Heinemann}
}

@article{loudet2004stokes,
  title={Stokes drag on a sphere in a nematic liquid crystal},
  author={Loudet, JC and Hanusse, P and Poulin, P},
  journal={Science},
  volume={306},
  number={5701},
  pages={1525--1525},
  year={2004},
  publisher={American Association for the Advancement of Science}
}

@article{ruhwandl1996friction,
  title={Friction drag on a particle moving in a nematic liquid crystal},
  author={Ruhwandl, RW and Terentjev, EM},
  journal={Physical Review E},
  volume={54},
  number={5},
  pages={5204},
  year={1996},
  publisher={APS}
}

@article{stark2001stokes,
  title={Stokes drag of spherical particles in a nematic environment at low Ericksen numbers},
  author={Stark, Holger and Ventzki, Dieter},
  journal={Physical Review E},
  volume={64},
  number={3},
  pages={031711},
  year={2001},
  publisher={APS}
}

@article{beer2025irreversibility,
  title={Irreversibility and symmetry breaking in the creation and annihilation of defects in active living matter},
  author={Beer, Avraham and Neimand, Efraim Dov and Corbett, Dom and Pearce, Daniel JG and Ariel, Gil and Yashunsky, Victor},
  journal={arXiv preprint arXiv:2508.15622},
  year={2025}
}

\setcounter{equation}{0}

\renewcommand{\theequation}{A\arabic{equation}}

\newpage

\onecolumngrid

\section*{Appendix}
\subsection{Active Flows of Defects}
\subsubsection{Perturbative Calculation}
\label{subsubsec:PerturbativeCalculation}
The equation determining the minimal energy defect morphology \eqref{eq:ELDefectDirector} can easily be solved perturbatively as a series in the elastic anisotropy $\kappa$. This was done to first order in \cite{ranganath1983energetics,hudson1989frank}; here we have need for the second order term, and so write
\begin{equation}
    \psi=(s-1)\theta+\kappa\frac{s(s-2)}{4(s-1)^2}\sin[2(s-1)\theta]+\kappa^2\frac{s(s-2)(5s^2-10s+4)}{64(s-1)^4}\sin[4(s-1)\theta]+O(\kappa^3).
    \label{eq:DirectorAnglePerturbative}
\end{equation}
The active force $\mathbf{f}=-\zeta\nabla\cdot(\mathbf{n}\mathbf{n})$ is then given by
\begin{align}
        \mathbf{f}^+&=-\zeta\left\lbrace\frac{1}{2r}\mathbf{e}_x+\frac{3\kappa}{4r}\mathbf{e}_r+\frac{\kappa^2}{32r}\left[3(-3+4\cos2\theta)\mathbf{e}_x+\frac{39}{2}\sin2\theta\mathbf{e}_y\right]\right\rbrace\label{eq:ActiveForcePlusHalfPerturbative},\\
        \mathbf{f}^-&=\zeta\left\lbrace\frac{1}{2r}(\cos2\theta\mathbf{e}_x-\sin2\theta\mathbf{e}_y)+\frac{5\kappa}{36r}\left[(\cos\theta+2\cos5\theta)\mathbf{e}_x+(\sin\theta-2\sin5\theta)\mathbf{e}_y\right]\right\rbrace,
        \label{eq:ActiveForceMinusHalfPerturbative}
\end{align}
for the $+1/2$ and $-1/2$ defects respectively. Note that for $+1/2$ defects the first order perturbation to the active force is a pure gradient, which is balanced by a corresponding pressure gradient, meaning the leading flow correction is at second order in $\kappa$. This is argued for on symmetry grounds in the main text.

The active flow may be found by integrating the active force against the Oseen tensor
\begin{equation}
        u_i(\mathbf{r})=\frac{1}{4\pi\mu}\int\text{d}^2\mathbf{r}'\left[\left(\text{log}\frac{\mathcal{L}}{|\mathbf{r}-\mathbf{r}'|}-1\right)\delta_{ij}+\frac{(\mathbf{r}-\mathbf{r}')_i(\mathbf{r}-\mathbf{r}')_j}{|\mathbf{r}-\mathbf{r}'|^2}\right]f_j(\mathbf{r}').
    \label{ActiveFlowIntegral}
\end{equation}
In evaluating these integrals we follow closely the approach of \cite{giomi2014defect}. Integrals arising from the first term in the Green's function take the form 
\begin{equation}        I_1=\int\text{d}^2\mathbf{r}'\left(\text{log}\frac{\mathcal{L}}{|\mathbf{r}-\mathbf{r}'|}-1\right)\mathbf{f}(\mathbf{r}')
\label{eq:I1General}
\end{equation}
and can be evaluated via the logarithmic expansion
\begin{equation}
    \log\frac{\mathcal{L}}{|\mathbf{r}-\mathbf{r}'|}=\log\frac{\mathcal{L}}{r_>}+\sum_{n=1}^{\infty}\frac{1}{n}\left(\frac{r_<}{r_>}\right)^n\cos[n(\theta-\theta')],
    \label{eq:TrigOrthogonality}
\end{equation}
where $r_>$ and $r_<$ denote $\text{max}(|\mathbf{r}|,|\mathbf{r}'|)$ and $\text{min}(|\mathbf{r}|,|\mathbf{r}'|)$ respectively, in conjunction with the orthogonality of trigonometric functions
\begin{equation}
    \int_0^{2\pi}\text{d}\theta'\cos[n(\theta-\theta')]\begin{cases}
    \cos m\theta' \\
    \sin m\theta'
    \end{cases}
    =\pi\delta_{nm}\begin{cases}
    \cos m\theta \\
    \sin m\theta
    \end{cases}.
\end{equation}
Integrals stemming from the second term have the form
\begin{align}
I_2 &= \int\mathrm{d}\mathbf{r}'\,
      \frac{(\mathbf{r}-\mathbf{r}')_i}{|\mathbf{r}-\mathbf{r}'|^2}
      \left[(x-x')f_x(\mathbf{r}')+(y-y')f_y(\mathbf{r}')\right]
      \nonumber \\
    &= 
      \begin{aligned}[t]
      &x\,\partial_{r_i}\!\int\!\mathrm{d}^2\mathbf{r}'\,\log|\mathbf{r}-\mathbf{r}'|\,f_x(\mathbf{r}')
        - \partial_{r_i}\!\int\!\mathrm{d}^2\mathbf{r}'\,\log|\mathbf{r}-\mathbf{r}'|\,x' f_x(\mathbf{r}')\\
      &+y\,\partial_{r_i}\!\int\!\mathrm{d}^2\mathbf{r}'\,\log|\mathbf{r}-\mathbf{r}'|\,f_y(\mathbf{r}')
        - \partial_{r_i}\!\int\!\mathrm{d}^2\mathbf{r}'\,\log|\mathbf{r}-\mathbf{r}'|\,y' f_y(\mathbf{r}')
      \end{aligned}
\label{eq:I2General}
\end{align}
and will be handled by utilising the relation
\begin{equation}
    \frac{(\mathbf{r}-\mathbf{r}')_i}{|\mathbf{r}-\mathbf{r}'|^2}=\partial_{r_i}\text{log}|\mathbf{r}-\mathbf{r}'|.
\end{equation}
Note that in all radial integrals the regions $r'<r$ and $r'>r$ must be treated separately.

We first consider the active flow associated with $+1/2$ defects. Upon substituting the active force given in \eqref{eq:ActiveForcePlusHalfPerturbative} the integrals in \eqref{eq:I1General} and \eqref{eq:I2General} evaluate to
\begin{align}
        I_1&=\left(1-\frac{9}{16}\kappa^2\right)\pi\left(r-R\log\frac{\mathcal{L}}{R}\right)\mathbf{e}_x+\frac{\pi\kappa^2r(3r-4R)}{16R}\left(\cos2\theta\mathbf{e}_x+\frac{13}{8}\sin2\theta\mathbf{e}_y\right),\\
        I_2&=\left(-\frac{\pi R}{2}-\frac{9\pi\kappa^2(7r^2-14rR+3R^2)}{128R}\right)\mathbf{e}_x-\frac{(16-9\kappa^2)\pi r}{48}(\cos2\theta\mathbf{e}_x+\sin2\theta\mathbf{e}_y)+\frac{\pi\kappa^2r}{64}(\cos4\theta\mathbf{e}_x+\sin4\theta\mathbf{e}_y).
\end{align}
Taking the combination $\frac{\zeta}{4\pi\mu}(I_1+I_2)$ and setting $\mathcal{L}=R\sqrt{e}$ yields the flow $\mathbf{u}^+$ given in \eqref{eq:ActiveFlowPlusHalf}.

Proceeding in the same way for $-1/2$ defects, the integrals in \eqref{eq:I1General} and \eqref{eq:I2General} evaluate to
\begin{align}
        I_1&=\frac{\pi r(3r-4R)}{12R}(\cos2\theta\mathbf{e}_x-\sin2\theta\mathbf{e}_y)+\frac{\pi\kappa r}{216R^4}\left[15R^4\left(2\log\frac{r}{R}-1\right)(\cos\theta\mathbf{e}_x+\sin\theta\mathbf{e}_y)+(3r^4-5R^4)(\cos5\theta\mathbf{e}_x-\sin5\theta\mathbf{e}_y)\right],\\
        I_2&=-\frac{\pi r}{15}(\cos4\theta\mathbf{e}_x+\sin4\theta\mathbf{e}_y)-\frac{5\pi\kappa r}{432}\left[6\left(2\log\frac{r}{R}-1\right)(\cos\theta\mathbf{e}_x+\sin\theta\mathbf{e}_y)+(\cos7\theta\mathbf{e}_x+\sin7\theta\mathbf{e}_y\right],
\end{align}
with the combination $\frac{\zeta}{4\pi\mu}(I_1+I_2)$ again providing the induced active flow $\mathbf{u}^-$, given in \eqref{eq:ActiveFlowMinusHalf}.

\subsubsection{Maximal Elastic Anisotropy}
We turn now to $+1/2$ defects with maximal elastic anisotropy. On account of the symmetries described in the main text the active flows are identical for both signs of elastic anisotropy. We therefore consider only the case $\kappa=1$ here, for which $\mathbf{f}=\frac{1}{r}\mathbf{e}_r$, provided $-\frac{\pi}{2}\leq\theta\leq\frac{\pi}{2}$. Outside this range the director is uniformly aligned and the active force vanishes. The principle difference compared to the prior perturbative calculation is that trigonometric functions are not orthogonal on this domain, meaning the orthogonality relations of \eqref{eq:TrigOrthogonality} must be replaced by
\begin{align}
        &\int_{-\pi/2}^{\pi/2}\cos(n(\theta-\theta'))\cos m\theta'\text{d}\theta'=
    \begin{cases}
        \frac{2\left[m\cos\frac{n\pi}{2}\sin\frac{m\pi}{2}-n\cos\frac{m\pi}{2}\sin\frac{n\pi}{2}\right]}{(m+n)(m-n)}\cos(n\theta) \qquad &m\neq n\\
        \frac{\pi}{2}\cos(n\theta) & m=n
    \end{cases},\\
        &\int_{-\pi/2}^{\pi/2}\cos(n(\theta-\theta'))\sin m\theta'\text{d}\theta'=
    \begin{cases}
        \frac{2\left[n\cos\frac{n\pi}{2}\sin\frac{m\pi}{2}-m\cos\frac{m\pi}{2}\sin\frac{n\pi}{2}\right]}{(m+n)(m-n)}\sin(n\theta) \qquad &m\neq n\\
        \frac{\pi}{2}\sin(n\theta) & m=n
    \end{cases}.
\end{align}
Apart from this the calculation is analogous to that of the preceding section, albeit lengthier, ultimately arriving at the active flow
\begin{equation}
        \begin{split}
            \mathbf{u}&=(3R-4r+\frac{r^2}{R})\mathbf{e}_x+\frac{4r}{3}(\cos2\theta\mathbf{e}_x+\sin2\theta\mathbf{e}_y)-\frac{r(r^3+4R^3)}{9R^3}(\cos2\theta\mathbf{e}_x-\sin2\theta\mathbf{e}_y)-\frac{r^2}{R}\sin2\theta\mathbf{e}_y\\
            &+\frac{r(5r^3-16R^3)}{180R^3}(\cos4\theta\mathbf{e}_x+\sin4\theta\mathbf{e}_y)+\frac{r(12r^5-25r^3R^2+16R^5)}{300R^5}(\cos4\theta\mathbf{e}_x-\sin4\theta\mathbf{e}_y)\\
            &+\sum_{k=3}^{\infty}\frac{(-1)^k}{2k-1}\left\lbrace\frac{1}{2k-1}\left[\frac{R}{2k}\left(\frac{r}{R}\right)^{2k}-\frac{4r}{2k+1}\right](\cos2k\theta\mathbf{e}_x+\sin2k\theta\mathbf{e}_y)\right.\\
            &\left.\qquad\qquad+\frac{1}{2k+1}\left[\frac{4r}{2k+1}-\frac{2k+1}{2k}R\left(\frac{r}{R}\right)^{2k}+\frac{2k-1}{2k+1}\frac{r^2}{R}\left(\frac{r}{R}\right)^{2k}\right](\cos2k\theta\mathbf{e}_x-\sin2k\theta\mathbf{e}_y)
            \right\rbrace
        \end{split},
        \label{eq:ActiveFlowMaximalAnisotropy}
\end{equation}
up to a factor of $\frac{\zeta}{4\pi\mu}$. This flow, truncated, is plotted in Figure \ref{fig:ActiveFlowPlusHalfDefectMaximalAnisotropy}.

\subsection{Elastic Energy of Defect Pairs}
Calculating the elastic energy of defect pairs is facilitated by expressing the director angle, $\phi$, in bipolar coordinates $(\eta,\rho)$, which are related to Cartesian coordinates by $x+\text{i}y=a\text{i}\cot\left(\frac{\eta+\text{i}\rho}{2}\right)$. This coordinate system has two foci, located at $(\pm a,0)$, with $\eta$ and $\rho$ at a given point having the respective geometrical interpretations of the angle subtended by the two foci and the logarithm of the ratio of the distances to the foci.

Bipolar coordinates, being singular at two points, naturally encode the director winding for a defect pair, with the defects located at the foci. For a pair of oppositely signed defects with elastic anisotropy
\begin{equation}
    \mathbf{n}=\cos\phi\hat{\mathbf{e}}_x+\sin\phi\hat{\mathbf{e}}_y, \qquad \phi=\frac{1}{2}\eta+\beta+O(\kappa).
\end{equation}
Here $\beta$ encodes the orientation of the defects, with the orientation vector $\mathbf{p}$ of the $+1/2$ defect as defined in \eqref{DefectOrientation} making an angle of $2\beta$ with the vector from the $-1/2$ to the $+1/2$ defect. The Frank free energy \eqref{FrankElasticAnisotropy} may then be expressed as
\begin{equation}\label{FrankBipolar}
    F=\frac{K_1+K_3}{4}\int\text{d}\rho\text{d}\eta\Omega^2\left[|\nabla\phi|^2+\kappa\left(|\nabla_{\mathbf{n}}\mathbf{n}|^2-|\nabla\cdot\mathbf{n}|^2\right)\right],
\end{equation}
where $\Omega^2$ is the scale factor for the area element, $\Omega=\frac{a}{\cosh\rho-\cos\eta}$.

In what follows we calculate the leading order contribution to the elastic energy in \eqref{FrankBipolar}. The first order change to the isotropic term in \eqref{FrankBipolar} comprises a total derivative and so we need only consider the zeroth order field in the anisotropic term. This anisotropic elasticity term is given by
\begin{align}
    &|\nabla_{\mathbf{n}}\mathbf{n}|^2-|\nabla\cdot\mathbf{n}|^2=-\frac{e^{-2\rho}}{32a^2z^3}\left\lbrace\cos2\beta\left[1-4e^{\rho}(z+z^5)+ \right.\left.e^{2\rho}(6+e^{2\rho})(z^2+z^4)-8e^{3\rho}z^3+z^6\right] \right. \\ \nonumber
& \qquad\qquad\qquad\qquad\qquad\qquad\qquad\left. +\text{i}\sin2\beta(z^2-1)\left[1-4e^{\rho}(z+z^3)+(1+6e^{2\rho}-e^{4\rho})z^2+z^4\right]\right\rbrace,
\end{align}
where we have used the substitution $z=e^{\text{i}\eta}$. Doing the same for the area element gives
\begin{equation}
\frac{K_1+K_3}{4}\kappa\Omega^2\text{d}\rho\text{d}\eta=-\frac{\text{i}a^2z(K_1+K_3)\kappa}{\left[(z-e^{-\rho})(z-e^{\rho})\right]^2}\text{d}\rho\text{d}z,
\end{equation}
so that in full the relevant integral becomes
\begin{equation}
    \begin{split}
    &\frac{K_1+K_3}{32}\kappa\int\text{d}\rho\oint_{|z|=1}\frac{e^{-2\rho}\text{d}z}{\left[z(z-e^{-\rho})(z-e^{\rho})\right]^2}\left\lbrace \text{i}\cos2\beta\left[1-4e^{\rho}(z+z^5)+e^{2\rho}(6+e^{2\rho})(z^2+z^4)-8e^{3\rho}z^3+z^6\right] \right. \\
    & \left.\qquad\qquad\qquad\qquad\qquad\qquad\qquad+\sin2\beta(1-z^2)\left[1-4e^{\rho}(z+z^3)+(1+6e^{2\rho}-e^{4\rho})z^2+z^4\right]\right\rbrace.
\label{eq:EnergyCorrectionOppositelySignedDefects}
\end{split}
\end{equation}
The residues and their locations are as follows
\begin{equation}
    \frac{K_1+K_3}{16}\kappa e^{-3\rho}(e^{2\rho}-1)\begin{cases} 
   -(\text{i}\cos2\beta+\sin2\beta) & z=0 \\
   \text{i}\cos2\beta+\sin2\beta & z=e^{\rho} \\
   -\text{i}\cos2\beta+\sin2\beta & z=e^{-\rho}
  \end{cases}.
\end{equation}
When $\rho<0$ the poles inside the integration contour are at $z=0$ and $z=e^{\rho}$ and so the integral vanishes, whereas it does not for $\rho>0$, leaving us with the following integral
\begin{align}
    &\frac{K_1+K_3}{4}\pi\kappa\cos2\beta\int_{0}^{\infty}\text{d}\rho e^{-3\rho}(e^{2\rho}-1)=\frac{K_3-K_1}{6}\pi\cos2\beta \\ \nonumber
    &=\frac{K_3-K_1}{6}\pi\frac{\mathbf{p}\cdot(\mathbf{r}_+-\mathbf{r}_-)}{|\mathbf{r}_+-\mathbf{r}_-|},
\end{align}
where $\mathbf{r}_+$ and $\mathbf{r}_-$ denote the positions of the two defects.

In the case of two $-1/2$ defects the calculation proceeds in the same fashion, with the $|\nabla\phi|^2$ contribution from the first order change in the director field still vanishing, and results in a torque of the same form but half the strength. Explicitly, \eqref{eq:EnergyCorrectionOppositelySignedDefects} becomes
\begin{align}
    &\frac{K_1+K_3}{16}\kappa\int\text{d}\rho\oint_{|z|=1}\frac{e^{-2\rho}\text{d}z}{z\left[(z-e^{-\rho})(z-e^{\rho})\right]^3}\left\lbrace \text{i}\cos2\beta\left[(z^2+z^4)\cosh\rho+2(z+z^5)\cosh2\rho-(1+z^6)\cosh3\rho-4z^3\right] \right. \\ \nonumber
    & \left.\qquad\qquad\qquad\qquad\qquad\qquad\qquad+\sin2\beta(1-z^2)\sinh\rho\left[1+z^4-4(z+z^3)\cosh\rho+2(1+z^2+z^4)\cosh2\rho\right]\right\rbrace,
\end{align}
such that the residues are given by
\begin{equation}
    \frac{K_1+K_3}{16}\kappa \begin{cases} 
   -\cos(2\beta-3\text{i}\rho) & z=0 \\
   -\text{i}\sinh3\rho~\text{e}^{-2\text{i}\beta} & z=e^{\rho} \\
   \text{i}\sinh3\rho~\text{e}^{2\text{i}\beta} & z=e^{-\rho}
  \end{cases}
\end{equation}
and the final integral becomes
\begin{equation}
        \frac{K_1+K_3}{8}\pi\kappa\cos2\beta\left(\int_{-\infty}^0\text{e}^{3\rho}\text{d}\rho+\int_0^{\infty}\text{e}^{-3\rho}\text{d}\rho\right)=\frac{K_3-K_1}{12}\pi\frac{\mathbf{p}\cdot(\mathbf{r}_+-\mathbf{r}_-)}{|\mathbf{r}_+-\mathbf{r}_-|}.
\end{equation}
\end{document}